\documentclass[11pt,utf8,a4paper]{article}
\pdfoutput=1 
\usepackage{braket,verbatim,bm,bbold,amsmath,amssymb,epsfig,float}
\usepackage{etoolbox}

\usepackage{jheppub} 
\usepackage[dvipsnames]{xcolor}  
\definecolor{linkcolor}{RGB}{50,100,130}  
\usepackage[]{caption}
\usepackage{subcaption}

\usepackage{tikz-cd,graphicx,hyperref}
\graphicspath{ {./figs/} }
\hypersetup{
colorlinks=true,    	
pdfstartview=FitV,
linkcolor= linkcolor,
citecolor= linkcolor,
urlcolor= linkcolor,
linktoc=page,
hyperindex=true,
hyperfigures=true,  
breaklinks=true
}
\makeatletter
\gdef\@fpheader{\vspace{0cm}}
\makeatother

\usepackage{tikz}  
\tikzset{
font={\fontsize{9pt}{12}\selectfont}}
\usepackage{pgfplots}
\usetikzlibrary{pgfplots.groupplots}
\pgfplotsset{compat=1.18}
\usetikzlibrary{arrows,shapes,positioning,calc,patterns}
\usetikzlibrary{decorations.markings,decorations.pathmorphing,decorations.pathreplacing,decorations.text}
\usepackage{mathtools}
\usepackage{epsfig,rotating,pifont}

\usepackage{empheq}

\def\be{\begin{equation}}
\def\ee{\end{equation}}
\def\bsub{\begin{subequations}}
\def\esub{\end{subequations}}
\def\ba#1\ea{\begin{align}#1\end{align}}
\newcommand{\beq}{\begin{equation}}
\newcommand{\eeq}{\end{equation}}

\def\Z{\mathcal{Z}}
\def\M{\mathcal{M}}
\def\H{\mathcal{H}}
\def\E{\mathcal{E}}

\def\tr{{\rm Tr}}
\def\Tr{{\rm Tr}}
\def\({\left(}
\def\){\right)}
\def\[{\left[}
\def\]{\right]}
\def\p{\partial}
\def\o{\omega}
\def\l{\lambda}
\def\a{\alpha}
\def\b{\beta}
\newcommand{\rom}[1]{\uppercase\expandafter{\romannumeral #1\relax}}
\def\e{\epsilon}
\def\D{\mathcal{D}}
\newcommand{\nc}{\newcommand}
\nc{\lb}{\label} 
\nc{\nn}{\nonumber} 
\nc{\ra}{\rangle} 
\nc{\la}{\langle} 
\nc{\Nn}{\mathcal{N}} 
\nc{\sigb}{\boldsymbol{\sigma_{\text{bd}}}}
\nc{\HA}{\mathcal{H}_A}
\nc{\HB}{\mathcal{H}_B}
\nc{\HC}{\mathcal{H}_C}
\nc{\ta}{\tilde{a}}
\nc{\Q}{\mathcal{Q}}
\nc{\T}{\mathcal{T}}
\nc{\Oo}{\mathcal{O}}
\nc{\re}{\mathrm{Re}}
\nc{\im}{\mathrm{Im}}
\nc{\scref}[1]{\ref{#1}} 
\renewcommand{\footnoterule}{\vfill\kern -3pt \hrule width 0.4\columnwidth \kern 2.6pt}

\definecolor{redCB}{RGB}{150,50,60}  
\definecolor{dteal}{RGB}{7,94,84}  
\definecolor{blueCB}{RGB}{108,144,170}

\makeatletter
\renewcommand*\env@matrix[1][\arraystretch]{%
\edef\arraystretch{#1}%
\hskip -\arraycolsep
\let\@ifnextchar\new@ifnextchar
\array{*\c@MaxMatrixCols c}}
\def\l@subsubsection#1#2{}
\makeatother

\title{Reflected entropy and Markov gap in Lifshitz theories}  
\author[a,b]{Cl\'ement Berthiere,}

\author[c,d,e]{Bin Chen,}
\author[c]{and Hongjie Chen}

\affiliation[a]{D\'epartement de Physique, Universit\'e de Montr\'eal, Montr\'eal, QC H3C 3J7, Canada}
\affiliation[b]{Centre de Recherches Math\'ematiques, Universit\'e de Montr\'eal, Montr\'eal, QC H3C 3J7, Canada}
\affiliation[c]{School of Physics and State Key Laboratory of Nuclear Physics and Technology,
Peking University, Beijing 100871, P.~R.~China}
\affiliation[d]{Collaborative Innovation Center of Quantum Matter, Beijing 100871, P.~R.~China}
\affiliation[e]{Center for High Energy Physics, Peking University, Beijing 100871, P.~R.~China}
\emailAdd{clement.berthiere@umontreal.ca}
\emailAdd{bchen01@pku.edu.cn}
\emailAdd{hongjie.chen@pku.edu.cn}

\abstract{
We study the reflected entropy in $(1+1)$--dimensional Lifshitz field theory whose groundstate is described by a quantum mechanical model. Starting from tripartite Lifshitz groundstates, both critical and gapped, we derive explicit formulas for the R\'enyi reflected entropies reduced to two adjacent or disjoint intervals, directly in \mbox{the continuum.} We show that the reflected entropy in Lifshitz theory does not satisfy monotonicity, in contrast to what is observed for free relativistic fields. We analytically compute the full reflected entanglement spectrum for two disjoint intervals, finding a discrete set of eigenvalues which is that of a thermal density matrix. Furthermore, we investigate the Markov gap, defined as the difference between reflected entropy and mutual information, and find it to be universal and nonvanishing, signaling irreducible tripartite entanglement in \mbox{Lifshitz} groundstates. We also obtain analytical results for the reflected entropies and the Markov gap in $2 + 1$ dimensions.
Finally, as a byproduct of our results on reflected entropy, we provide exact \mbox{formulas} for two other entanglement-related quantities, namely the computable cross-norm negativity and the operator entanglement entropy.
}

\begin{document} 

\maketitle
\flushbottom

\bigskip

\section{Introduction}

Quantum information has revolutionized the study of quantum many-body systems \cite{Amico:2007ag,Calabrese:2009qy,Laflorencie:2015eck}, offering valuable insights into their universal properties \cite{Vidal:2002rm,Calabrese:2004eu,2006PhRvL..96j0603L,Eisert:2008ur,Bueno:2015rda,Fursaev:2016inw,Casini:2016fgb,Berthiere:2018ouo,Berthiere:2019lks} and revealing hidden phases that are otherwise undetectable using local probes \cite{Kitaev:2005dm,2006PhRvL..96k0405L,2015arXiv150802595Z}. 
Central to these developments is the concept of entanglement between two parties in a pure state. 
Much of the focus has thus revolved around the entanglement entropy, the measure that captures the amount of EPR entanglement present asymptotically in a pure state, which is the sole form of bipartite pure-state entanglement.
In contrast, in systems shared by three or more parties, several inequivalent classes of entanglement can be identified and no single measure captures all its aspects. Although the entanglement entropy has provided crucial understanding in condensed-matter physics, quantum field theory, and holography, it cannot distinguish among the various forms of multipartite entanglement. Furthermore, the entanglement entropy of subregions is ill-defined in continuum theories, thus requiring careful UV regularization. To overcome these challenges, we aim to investigate other information-theoretic quantities that can give a more complete picture of the structure of entanglement, and which remain well-defined in the continuum limit.
One such quantity is the reflected entropy, defined for bipartite \textit{mixed} states \cite{Dutta:2019gen}.

Since its introduction, the reflected entropy has attracted much attention in various contexts, including holographic models \cite{Jeong:2019xdr,Bao:2019zqc,Chandrasekaran:2020qtn,Basu:2021awn,Li:2020ceg,Li:2021dmf,Ling:2021vxe,Akers:2022max,Basu:2022nds,Basak:2022cjs,Chen:2022fte,Vasli:2022kfu,Lu:2022cgq,Afrasiar:2022fid,Afrasiar:2023jrj}, free theories \cite{Bueno:2020vnx,Bueno:2020fle,Dutta:2022kge,Basak:2023uix}, topological systems \cite{Berthiere:2020ihq,Liu:2021ctk,Sohal:2023hst}, conformal field theory (CFT) \cite{Kusuki:2019rbk,Kudler-Flam:2020url,Moosa:2020vcs,Kudler-Flam:2020xqu,Berthiere:2023gkx} or random tensor networks \cite{Akers:2021pvd,Akers:2022zxr}.
In quantum field theory, the reflected entropy of two disjoint regions is the von Neumann entropy \cite{Longo:2019pjj} of a canonically defined type \rom{1} algebra \cite{Doplicher:1984zz}, and is as such well-defined (see, e.g., \cite{Longo:2019pjj,Bueno:2020vnx} for more details).
Although not a genuine bipartite correlation measure \cite{Hayden:2023yij}, the reflected entropy is a probe of tripartite entanglement \cite{Akers:2019gcv,Hayden:2021gno}. Indeed, the Markov gap, defined as the difference between the reflected entropy and mutual information, has been shown to detect beyond bipartite and GHZ-type entanglements \cite{Zou:2020bly}.
To fully characterize the entanglement structure of a system, it is crucial to investigate its multipartite entanglement, see \cite{Umemoto:2018jpc,Akers:2019gcv,Zou:2020bly,Hayden:2021gno,Liu:2021ctk,Agon:2021lus,tam2022topological,parez2022multipartite,liu2023multipartite,carollo2022entangled,parez2022analytical,maric2022universality} for recent progress.

\smallskip
Here, we study the reflected entropy for Lifshitz groundstates. Quantum Lifshitz field theories constitute rare examples of non-relativistic theories which admit analytic treatment. At their critical points, they exhibit anisotropic scaling between space and time
\be\label{LFTscaling}
t\rightarrow \lambda^z t\,,\quad\; x\rightarrow \lambda x\,,\quad\;\lambda>0\,,
\ee
with dynamical exponent $z\neq1$. 
Such theories are also relevant in the context of condensed matter physics. For instance, the $z=2$ Lifshitz scalar field in $2+1$ dimensions describes the critical point of the quantum dimer model in the continuum limit  \cite{Rokhsar:1988zz,Ardonne:2003wa}. 
Bipartite entanglement properties of Lifshitz theories have been extensively studied, see, e.g.,
\cite{Fradkin:2006mb,Hsu:2008af,PhysRevB.80.184421,Oshikawa:2010kv,2011PhRvL.107b0402Z,Zhou:2016ykv,Chen:2016kjp,chen2017quantum,MohammadiMozaffar:2017nri,MohammadiMozaffar:2017chk,Angel-Ramelli:2019nji,Angel-Ramelli:2020wfo,Angel-Ramelli:2020xvd}. 
Lifshitz groundstates belong to the more general class of Rokhsar-Kivelson states (see, e.g., \cite{Rokhsar:1988zz,Ardonne:2003wa,henley2004classical,2005AnPhy.318..316C}), 
which have recently been shown to be separable for two (or more) disconnected regions \cite{Boudreault:2021pgj,Parez:2022ind}--a fundamental feature of the entanglement structure of such quantum states. 

In the present work, we compute the reflected entropy and Markov gap for Lifshitz groundstates composed of three parties, with the goal of obtaining a better understanding of tripartite entanglement in such theories.
Note that the reflected entropy is related to two other information-theoretic quantities, namely the operator entanglement entropy \cite{Zanardi:2001zza,2007PhRvA..76c2316P,Dubail:2016xht} and the computable cross-norm or realignment (CCNR) negativity \cite{Yin:2022toc}, both of which shall be investigated as well.
In what follows, we review  the definition of (Rényi) reflected entropy and we present in more details the Lifshitz theories we shall focus on.

\pagebreak
\noindent\textbf{Definition of reflected entropy and replica formulation}
\smallskip

\noindent Consider a general mixed bipartite quantum state $\rho_{AB}$ on a finite Hilbert space $\H_A\otimes \H_B$. The state can be canonically purified as the pure state $\ket{\sqrt{\rho_{AB}}}$ in a doubled Hilbert space $(\H_A\otimes \H_B)\otimes(\H_{A^*}\otimes \H_{B^*})$. The reflected entropy $S^R$ is defined as von Neumann (entanglement) entropy associated to the reflected density matrix $\rho_{AA^*} = \tr_{BB^*}(\ket{\sqrt{\rho_{AB}}}\bra{\sqrt{\rho_{AB}}})$, i.e.~$S^R(A:B)=-\tr\big(\rho_{AA^*}\log\rho_{AA^*}\big)$. It is expected that working on a lattice and taking the continuum limit, this definition coincides with that using von-Neumann algebras \cite{Longo:2019pjj}. 

A replica formulation of reflected entropy provides a convenient tool for computations. It involves two replica indices, $m$ and $n$. The latter is the standard R\'enyi index, while the former is related to the (generalized) purification $\ket{\rho_{AB}^{m/2}}$ of $\rho_{AB}^m$ with \mbox{$m\in2\mathbb{Z}^+$}, such that $\rho_{AB}^m=\tr_{A^*B^*}\big(\ket{\rho_{AB}^{m/2}}\bra{\rho_{AB}^{m/2}}\big)$. One then defines the reflected density matrix $\rho_{AA^*}^{(m)}=\tr_{BB^*}\big(\ket{\rho_{AB}^{m/2}}\bra{\rho_{AB}^{m/2}}\big)$, and introduces the $(m,n)$--Rényi reflected entropy as 
\be\label{eq:REmn}
S^{R}_{m,n}(A:B)=\frac{1}{1-n}\log\frac{\Tr\big(\rho_{AA^*}^{(m)}\big)^n}{\big(\tr\rho_{AB}^m\big)^n}\,,
\ee
where we used $\tr\rho_{AA^*}^{(m)}=\tr\rho_{AB}^m$ for normalization. 
The $m$--reflected entropy $S^R_{m}(A:B) = \lim_{n\rightarrow1} S^R_{m,n}(A:B)$ is obtained by analytic continuation $n\rightarrow1$, and the original reflected entropy $S^R(A:B)$ is recovered by further taking the limit $m\rightarrow 1$. Note that if $\rho_{AB}$ is pure, the reflected entropy is twice the entanglement entropy associated to the bipartition $A/B$, i.e.~$S^R_{m,n}(A:B)=2S_n(A)$ where $S_n(A)=\frac{1}{1-n}\log\frac{\tr\rho_A^n}{(\tr\rho_A)^n}$. The reflected entropy enjoys many interesting properties, see \cite{Dutta:2019gen}.

\medskip\smallskip
\noindent \textbf{Lifshitz scalar field theory and Rokhsar-Kivelson groundstates}
\smallskip

\noindent In this paper, we are interested in a certain class of nonrelativistic quantum field theories, called Lifshitz theories, which possess the remarkable feature that their groundstate wavefunctional takes a local form, given in terms of the action of a classical model---a Rokhsar-Kivelson wavefunctional \cite{Rokhsar:1988zz,henley2004classical}.

We shall consider both the (noncompact) $z=2$ Lifshitz critical boson and its massive deformation \cite{Boudreault:2021pgj} in $1+1$ dimensions, described by the Hamiltonian 
\beq\label{E:H_Lifshitz}
H=\frac{1}{2}\int d x \left(\Pi^2 + (\p_x^2 \phi)^2 + 2 \omega^2 (\p_x \phi)^2 + \omega^4 \phi^2 \right),
\eeq
with canonical commutation relations $[\phi(x),\Pi(x')]=i\delta(x-x')$, and $\Pi(x)=-i\delta / \delta\phi(x)$ in the Schrödinger picture.
The groundstate can be expressed in terms of a path integral of a one-dimensional Euclidean theory
\beq\label{E:Psi_0}
\ket{\Psi_0}=\frac{1}{\sqrt{\mathcal{Z}}}\int \D \phi \,e^{-\frac{1}{2}S_{\rm cl}[\phi]}\ket{\phi}, \qquad S_{\rm cl}[\phi]=\int dx \big((\p_x\phi)^2+\omega^2\phi^2\big)\,,
\eeq
with normalization factor $\mathcal{Z}=\int \mathcal{D}\phi\, e^{-S_{\rm cl}[\phi]}$.
In one dimension, $\mathcal{Z}$ corresponds to the partition function of a single particle with classical action $S_{\rm cl}[\phi]$, i.e.~an (Euclidean) harmonic oscillator of ``mass" $M=2$ and ``frequency" $\omega$. In the critical massless case, $\omega=0$, the theory is invariant under Lifshitz scaling \eqref{LFTscaling} with $z=2$, and the classical action is that of a free nonrelativistic particle.
The propagator of the Euclidean harmonic oscillator, defined as $K(\phi, \phi';\ell)=\int_{\phi(0)=\phi}^{\phi(\ell)=\phi'}\D\phi\, e^{-S_{\rm cl}[\phi]}$, reads
\beq\label{E:propagator_mass}
\begin{aligned}
K(\phi, \phi';\ell) =\sqrt{\frac{\omega}{\pi  \sinh \omega \ell}}\exp\bigg[
					\frac{-\omega\big((\phi^2 + {\phi'}^2)\cosh \omega\ell - 2\phi\phi'\big)}{\sinh\omega \ell}\bigg],
\end{aligned}
\eeq 
which in the massless limit $\omega=0$ reduces to 
\beq\label{E:propagator_Lif}
K(\phi, \phi';\ell)  = \frac{1}{\sqrt{\pi\ell}}e^{-\frac{(\phi'-\phi)^2}{\ell}}.
\eeq 
Vacuum expectation values, associated to $\ket{\Psi_0}$, of local operators can then be expressed in terms of the above propagator.

\medskip\smallskip
\noindent\textbf{Tripartition and reduced density matrix}
\smallskip

\noindent The density matrix operator corresponding to the Lifshitz groundstate $\ket{\Psi_0}$ is
\beq\label{E:rho}
\rho=\frac{1}{\mathcal{Z}}\int\mathcal{D}\phi\mathcal{D}\phi' \, e^{-(S_{\rm cl}[\phi]+S_{\rm cl}[\phi']) /2}|\phi\rangle\langle\phi'| \, ,
\eeq
with implicit boundary conditions (BCs) on the full system (e.g.~Dirichlet, Neumann or periodic). In this work, we consider a class of bipartite mixed states obtained by tracing out the degrees of freedom of one of the subsystems in a tripartite pure state, which we choose to be the groundstate $\ket{\Psi_0}$ in \eqref{E:Psi_0}. More precisely, beginning with $\ket{\Psi_0}$, we divide our system into three non-overlapping subsystems, $A$, $B$ and $C$, and consider the reduced density matrix on $A\cup B$, which is in general that of a mixed state.
The reduced density matrix on $A\cup B$ reads 
\ba
\rho_{AB}=\hspace{-3pt}\int_{\rm JC}\hspace{-3pt} \scalebox{0.97}{$\D\phi_A \D\phi'_{\hspace*{-1pt}A} \D\phi_B \D\phi'_B \D\phi_C\,$}e^{-\frac12\left(\hspace{-1pt}S_{\rm cl}[\phi_{\hspace*{-1pt}A}]+S_{\rm cl}[\phi'_{\hspace*{-1pt}A}]+S_{\rm cl}[\phi_{B}]+S_{\rm cl}[\phi'_{\hspace*{-1pt}B}]+2S_{\rm cl}[\phi_C]\hspace{-1pt}\right)}\hspace{-2pt}\ket{\phi_A}\hspace{-2.pt}\ket{\phi_B}\hspace{-2pt}\bra{\phi'_{\hspace*{-.75pt}A}}\hspace{-2pt}\bra{\phi'_B}\hspace{-1.5pt},
\ea
with junction conditions at each interface between $A$, $B$, and $C$ depending on the geometry. Here, $\phi_A$ denotes fields defined on $A$, and similarly for $\phi_B$ and $\phi_C$. For example, for $A,B$ disjoint as in Figure \scref{fig:disjoint-l}, the junction conditions at the interfaces $\Gamma_X$ between $X=\{A,B\}$ and $C$ are $\phi_X|_{_{\Gamma_X}}\hspace{-3pt}=\phi'_X|_{_{\Gamma_X}}\hspace{-3pt}=\phi_C|_{_{\Gamma_X}}$.
We may leave $\rho_{AB}$ unnormalized, such that $\tr \,\rho_{AB}=\Z$ is the partition function of the underlying classical model on a \emph{graph} that encodes the boundary and junction conditions.

The reflected density matrix $\rho_{AA^*}^{(m)}$ can be obtained by first computing $\rho_{AB}^{m/2}$ with $m\in2\mathbb{Z}^+$. A canonical doubling of the Hilbert space provides the simplest purification $\ket{\rho_{AB}^{m/2}}$, that is we turn bras of $\rho_{AB}^{m/2}$ into kets in $\H_{A^*}\otimes \H_{B^*}$. The moments $\tr\big(\rho_{AA^*}^{(m)}\big)^n$ can then be calculated as the partition function $\Z_{m,n}$ on a replica graph $\M_{m,n}$,
\ba
\Z_{m,n} = \int_{\M_{m,n}}\D\phi \,e^{-S_{\rm cl}[\phi]} = \tr\big(\rho_{AA^*}^{(m)}\big)^n \,,
\ea
such that, using \eqref{eq:REmn}, the $(m,n)$--Rényi reflected entropy follows as
\ba\label{eq:REmnZ}
S^R_{m,n}(A:B)=\frac{1}{1-n}\log\frac{\Z_{m,n}}{\(\Z_{m,1}\)^n}\,.
\ea

\medskip\smallskip
\noindent\textbf{Organization of the paper}
\smallskip

\noindent In Section~\ref{sec:11}, we compute the reflected entropy reduced to two adjacent or disjoint intervals from tripartite Lifshitz groundstates, both critical and massive. For two disjoint regions, the reflected entropy is finite, and the associated spectrum is discrete. For two adjacent regions, it is nonmonotonic under partial trace. We study the Markov gap in Section~\ref{sec:mi}, which is found to be positive, signaling nontrivial tripartite entanglement in Lifshitz groundstates.
We discuss the CCNR negativity and the operator entanglement entropy in Section~\ref{sec:oe}. An incursion is made in $2+1$ dimensions in Section \ref{sec:21}, where we obtain analytic results for the reflected entropy in certain limits. 
We conclude in Section~\ref{conclu} with a summary of our main results, and give an outlook on future study. Finally, some technical details and further calculations can be found in the three appendices that complete this work.

\section{Reflected entropies}
\label{sec:11}

\noindent In this section, we study the R\'enyi reflected entropies for non-compact Lifshitz theories in $1+1$ dimensions.\footnote{\flushbottom A ``modular flowed"  version of the reflected entropy, called the deflected entropy, can also be computed analytically for Lifshitz theories. We discuss the deflected entropy  of two disjoint regions in Appendix~\ref{deflected}.} We consider the critical (massless) Lifshitz scalar and its massive deformation introduced in \cite{Boudreault:2021pgj}. Extending the method developed in \cite{chen2017quantum,chen2017gapless,Boudreault:2021pgj}, we construct the reflected density matrix for bipartite mixed states corresponding to two adjacent or disjoint subsystems, and compute the reflected entropies for arbitrary R\'enyi indices $m, n$.
As a byproduct of our results, we find that the reflected entropy for Lifshitz theories is not monotonically nonincreasing under partial trace, contrarily to what is observed for free relativistic fields \cite{Bueno:2020fle}. We also compute the spectrum of the reflected density matrix for two disjoint subsystems, finding a discrete set of eigenvalues.

\subsection{Critical (massless) Lifshitz scalar}

We study the reflected entropies for the critical Lifshitz scalar defined by the Hamiltonian \eqref{E:H_Lifshitz} setting the mass $\omega=0$. The corresponding groundstate \eqref{E:Psi_0} with $\omega=0$ encodes the partition function of a free nonrelativistic particle whose propagator is given in \eqref{E:propagator_Lif}.

\begin{figure}[H]
\centering
\includegraphics[scale=1.1]{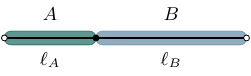}
\caption{Bipartite configuration of a finite system with Dirichlet BCs: two complementary regions $A, B$ of respective length $\ell_A, \ell_B$. Small hollow circles denotes Dirichlet BCs at the end of the system, while a black solid dot indicates that the value of $\phi$ is free.}
\label{fig:bipartite}
\end{figure}

\begin{figure}[t]
\centering
\includegraphics[scale=1.]{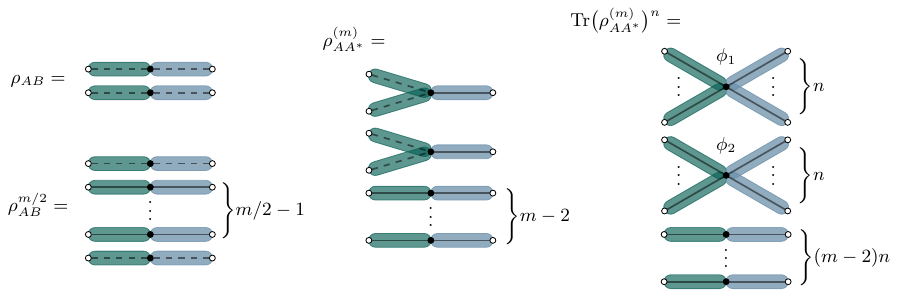}
\caption{Graphical representation of the calculations leading to the reflected entropy in the bipartite (pure state) case. We use different colors to differentiate lengths $\ell_A$ (green) and $\ell_B$ (blue). Dashed lines denote the ``indices'' of the matrix, while solid lines mean that trace has been taken.  Each value of the field at the interface between $A$ and $B$, denoted $\phi_i$, needs to be integrated over.}
\label{fig:bipartite-replica}
\end{figure}

To illustrate the purification process and the replica method for the reflected entropy outlined in the introduction, let us first consider the bipartite case for which \mbox{$\rho_{AB}=\ket{\Psi_0}\hspace{-2pt}\bra{\Psi_0}$} is pure. We work on a finite one-dimensional system with Dirichlet BCs at both ends, as depicted in Figure \ref{fig:bipartite}. The partition function \mbox{$\Z_{m,n}\hspace{-1pt}\equiv\tr\big(\rho_{AA^*}^{(m)}\big)^n\hspace{-1pt}$ of the free particle on the} (disconnected) replica graph $\M_{m,n}$ can be calculated using the procedure shown in Figure~\ref{fig:bipartite-replica}.

First, we compute $\rho_{AB}^{m/2}$, which is simply given by $\rho_{AB}^{m/2}=\Z^{m/2-1}\rho_{AB}$. Second, taking two copies of $\rho_{AB}^{m/2}$ treated as bra and ket, respectively, and tracing over $BB^*$ results in the reflected density matrix $\rho_{AA^*}^{(m)}$\,, which is proportional to $\rho_A\otimes\rho_A$\,. Finally, $\big(\rho_{AA^*}^{(m)}\big)^n$ is proportional to $\rho_A^n\otimes\rho_A^n$. In Figure \ref{fig:bipartite-replica}, we use dashed lines for bras and kets, and solid lines to represent traces taken, which are simply the propagators.
We find for $\rho_{AB}$ pure,
\ba
\Z_{m,n}& =\Z^{(m-2)n}\int d \phi_1 d \phi_2\,K(0,\phi_1;\ell_A)^nK(0,\phi_1;\ell_B)^nK(0,\phi_2;\ell_A)^nK(0,\phi_2;\ell_B)^n\nonumber\\
 &=\Z^{(m-2)n}\(\pi^2\ell_A\ell_B\)^{-n}\int d\phi_1 d\phi_2\exp\[-\frac {n\phi_1^2}{\ell_A}-\frac  {n\phi_1^2}{\ell_B}-\frac {n\phi_2^2}{\ell_A}-\frac  {n\phi_2^2}{\ell_B}\]\nonumber\\
 &=\Z^{(m-2)n}\(\pi^2\ell_A\ell_B\)^{-n}\frac\pi n\frac{\ell_A\ell_B}{\ell_{AB}}\,,\label{eq:bipartite-z}
\ea
where $\ell_{AB}=\ell_{A}+\ell_B$. The $(m,n)$--Rényi reflected entropy follows using \eqref{eq:REmnZ},
\bsub
\ba
\label{eq:renyi-re-bipartite}
\begin{split}
S^R_{m,n}(A:B)=\log\frac{\ell_A\ell_B}{\ell_{AB}\e}+ \gamma_n\,,
\end{split}
\ea
with nonuniversal constant term $\gamma_n=\frac{1}{n-1}\log n +\log\pi$. Taking the limit $n\rightarrow1$ we find
\be
\label{eq:re-bipartite}
S^R_{m}(A:B)=\log\frac{\ell_A\ell_B}{\ell_{AB}\e}+\gamma\,,
\ee
\esub
with $\gamma\equiv \gamma_1$. Note that we have reinstated a UV cutoff $\e$; whenever there is an unequal number of $\ell_i$ in the numerator and denominator in a logarithm, it is understood that we should add the corresponding power of cutoff length $\e$. This is due to the fact that the field $\phi$ has mass dimension $-1/2$, which we have not taken into account in the path integral. 
As expected, the reflected entropy is equal to twice the Rényi entanglement entropy \cite{chen2017quantum,Angel-Ramelli:2020wfo}
\be
\label{eq:ee-bipartite}
S_n(A)=\frac 1 2 \log\frac{\ell_A\ell_B}{\ell_{AB}\e}+\frac{\gamma_n}2\,,
\ee
which could already be realized from Figure \ref{fig:bipartite-replica}.

\subsubsection{Disjoint regions on an interval}\label{sec:disjoint}

\begin{figure}[t]
\centering
\begin{subfigure}[t]{0.4\textwidth}
\centering
\includegraphics[scale=1.05]{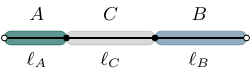}
\caption{}
\label{fig:disjoint-l}
\end{subfigure}
\begin{subfigure}[t]{0.55\textwidth}
\centering
\includegraphics[scale=1.05]{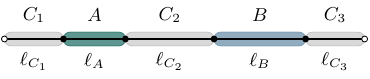}
\caption{}
\label{fig:disjoint-r}
\end{subfigure}
\caption{Two disjoint regions $A$ and $B$ on a finite system with Dirichlet BCs.}
\label{fig:disjoint}
\end{figure}

We now turn to the more interesting case of two disjoint subsystems on an interval with Dirichlet BCs. Two such configurations are shown in Figure \ref{fig:disjoint}. As mentioned in the Introduction, the reduced density matrix of two disjoint subsystems is separable, implying that they are not entangled, and correlations being classical and quantum non-entangling.

\begin{figure}[b]
\centering
\includegraphics[scale=1.02]{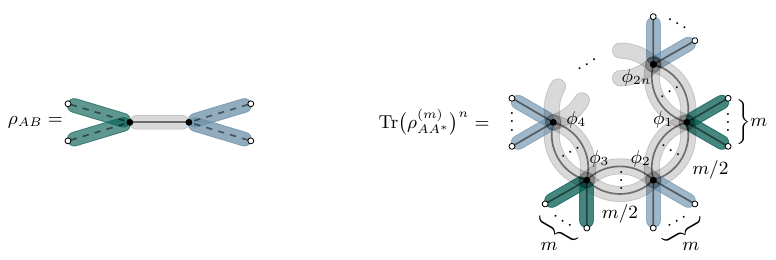}
\caption{Reduced density matrix $\rho_{AB}$ (left) and replica graph $\mathcal{M}_{m,n}$ (right) for two disjoint regions as in Figure \scref{fig:disjoint-l}. There are $m/2$ copies of length $\ell_C$ edges between two vertices labeled by $\phi_i$ and $\phi_{i+1}$.} 
\label{fig:disjoint-replica}
\end{figure}

We first consider the configuration depicted in Figure \scref{fig:disjoint-l}, where $A$ and $B$ are adjacent to the boundaries of the system. The replica graph $\M_{m,n}$ is drawn in Figure~\ref{fig:disjoint-replica}, and the corresponding partition function reads

\ba
\Z_{m,n}&= \int d\phi_1\cdots d\phi_{2n}\,K(0,\phi_1;\ell_A)^m K(\phi_1,\phi_2;\ell_C)^{m/2}K(0,\phi_2;\ell_B)^mK(\phi_2,\phi_3;\ell_C)^{m/2}\nn\\
&\hspace{1.9cm}\times K(0,\phi_3;\ell_A)^{m} \cdots K(\phi_{2n-1},\phi_{2n};\ell_C)^{m/2}K(0,\phi_{2n};\ell_B)^m K(\phi_{2n},\phi_{1};\ell_C)^{m/2}\nn\\
&= \pi^{(1-3m/2)n}(\ell_A\ell_B\ell_C)^{-mn/2}\( \det M_{m,n}\)^{-1/2}, \label{eq:disjoint-partition}
\ea
where $M_{m,n}$ is a $2n\times 2n$ matrix given in \eqref{eq:disjoint-matrix}.
Let us define the cross-ratio $\eta$ as 
\ba
\eta=\frac{\ell_A\ell_B}{\ell_{AC}\ell_{BC}}\,.
\ea
Using formula \eqref{eq:det1} in Appendix \ref{app:det}, we find
\be\label{eq:det_disc_D}
\det M_{m,n}=\(\frac{m}{2\ell_C}\)^{2n} 4\sinh^2\hspace{-2pt}\(n~ \mathrm{arccosh}\,\sqrt{\eta^{-1}}\)\,
\ee\pagebreak

\noindent which provides a nice analytic continuation in $n$. Plugging \eqref{eq:disjoint-partition} together with \eqref{eq:det_disc_D} into \eqref{eq:REmnZ}, the R\'enyi reflected entropies read
\ba
S^R_{m,n}(A:B)
=\frac{1}{n-1}\log\frac{\big(\sqrt{1-\eta}+1\big)^{2n}-\eta^{n}}{\big(\big(\sqrt{1-\eta}+1\big)^{2}-\eta\big)^n}\,. \label{eq:renyi-re}
\ea
For two disjoint subsystems, we find that the reflected entropies are finite and, interestingly, independent of $m$. In the limit $n\rightarrow1$ we obtain
\be
\label{eq:re}
S^R_{m}(A:B)=\frac{1}{\sqrt{1-\eta}}\log\(\frac{1+\sqrt{1-\eta}}{\sqrt\eta}\)  -\log \(2\sqrt{\eta^{-1}-1}\).
\ee

For the more general configurations where the endpoints of $A$ and $B$ are not boundaries of the system (see Figure \ref{fig:disjoint-r}), the reflected entropies are given by \eqref{eq:renyi-re} and \eqref{eq:re}, with cross-ratio 
\ba
\eta=\frac{\ell_{AC_1}\ell_{BC_3}}{\ell_{C_1AC_2}\ell_{C_2BC_3}}\,.
\ea

\medskip
\noindent \textbf{Taking limits:\,}
As a function of $\eta\in[0,1]$, $S^R_m(A:B)$ is monotonically increasing; it vanishes for $\eta=0$ and diverges for $\eta=1$.
For two disjoint intervals adjacent to the boundaries of the system, as in Figure \scref{fig:disjoint-l}, in the limit $\ell_C\gg \ell_A,\,\ell_B$ ($\eta\rightarrow0$) of two widely separated regions, the reflected entropies behave as
\be
\label{eq:far}
S^R_{m}(A:B)\sim -\frac{1}{4}\eta \log \eta +O(\eta)\,.
\ee
In the opposite limit of small separation $\ell_C \rightarrow 0$ ($\eta\rightarrow1$), the reflected entropies have asymptotics
\be
\begin{aligned}
\label{eq:near}
S^R_{m}(A:B) &\simeq -\frac 1 2 \log (1-\eta)+1-\log 2\\
&= \frac 1 2\log \frac{\ell_A\ell_B}{\ell_{AB}\ell_C}+1-\log 2\,.
\end{aligned}
\ee
Identifying $\ell_C\sim\epsilon$, we find that expression \eqref{eq:near} does not reproduce the bipartite \mbox{result} \eqref{eq:re-bipartite}, but only half of it. This suggests that the reflected entropy is not a good regularization of entanglement entropy, similarly as the mutual information (see Section \ref{sec:mi}).

In the configuration of Figure \scref{fig:disjoint-r}, letting $\ell_{C_1}, \ell_{C_3}\rightarrow\infty$ while keeping $\ell_A, \ell_B$ and $\ell_{C_2}$ finite, i.e $\eta\rightarrow 1$, the reflected entropies diverge. We show in Appendix \ref{app:neu} that letting $\ell_{C_1}, \ell_{C_3}\rightarrow\infty$ is equivalent to imposing Neumann BCs on the endpoints. The divergence is thus due to the presence of a zero mode.

\pagebreak
\subsubsection{Adjacent regions on an interval}
\label{sec:adj-interval}

\begin{figure}[t]
\centering
\includegraphics[scale=1.1]{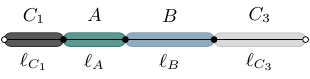}
\caption{Two adjacent regions $A$ and $B$ on finite system with Dirichlet BCs.}
\label{fig:adjacent}
\end{figure}

\begin{figure}[b]
\centering
\includegraphics[scale=1.05]{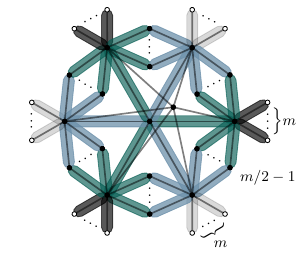}
\caption{Replica graph $\mathcal{M}_{m,3}$ for two adjacent regions as in Figure~\scref{fig:adjacent}. The vertices of degree two (between blue and green edges) can be neglected due to the property of propagators $\int d\phi_2 K(\phi_1,\phi_2;\ell_1)K(\phi_2,\phi_3;\ell_2)=K(\phi_1,\phi_3;\ell_1+\ell_2)$. Apart from the boundary vertices (hollow circles), we have $2n+2$ vertices: $2n$ of them are on a circle-like shape, where adjacent ones are connected by $m/2-1$ edges of length $\ell_A+\ell_B$; the other two vertices connect to each of these (for clarity, we color the edges of one of the two only). When $m=2$, the circle vanishes. When $\ell_{C_1}=0$ or $\ell_{C_3}=0$, the circular shape breaks down, due to Dirichlet BCs. When $\ell_{C_1}=\ell_{C_3}=0$, we recover Figure \ref{fig:bipartite-replica}.}
\label{fig:adjacent-replica}
\end{figure}

We now consider the reflected entropy of two adjacent regions on an interval, as depicted in Figure \scref{fig:adjacent}. We draw the replica graph $\M_{m,3}$ in Figure \ref{fig:adjacent-replica}, which can easily be generalized to $\M_{m,n}$.
The corresponding partition function reads
\be
\Z_{m,n}=\pi^{1-3mn/2}\ell_A^{-n}\ell_B^{-n}\ell_{C_1}^{-mn/2}\ell_{C_3}^{-mn/2}(\ell_A+\ell_B)^{-(m-2)n/2}\( \det M_{m,n}\)^{-1/2},
\ee
where $M_{m,n}$ is a $(2n+2)\times (2n+2)$ matrix given in \eqref{eq:adjacent-matrix}.
Up to a total factor $\frac{m/2-1}{\ell_A+\ell_B}$, the matrix $M_{m,n}$ is of the form \eqref{eq:m4}. Defining
\be
\label{eq:adjacent-t}
 u=\frac{(m-2)^2\ell_A\ell_B\ell_{C_1}\ell_{C_3}}{\(2\ell_A\ell_{C_3}+m\ell_B\ell_{ABC_3}\)\(2\ell_B\ell_{C_1}+m\ell_A\ell_{ABC_1}\)}\,,
\ee
the reflected entropies of two adjacent subsystems are found to be
\bsub
\ba
S^R_{m,n}(A:B)=\frac 1 2 \log \frac{(m-2)^2\ell_A^2\ell_B^2\ell_{C_1}\ell_{C_3}}{4m^2\ell_{AB}^3\ell \e^2}+\frac{1}{n-1}\log\frac{\big(1+\sqrt{1-u}\big)^{2n}-u^n}{2u^{(n-1)/2}\sqrt{1-u}\big(1+\sqrt{1-u}\big)^n}
+\gamma_n \,. \label{eq:renyi-re-adjacent-2}
\ea
\be
S^R_m(A:B)=\frac 1 2 \log \frac{(m-2)^2\ell_A^2\ell_B^2\ell_{C_1}\ell_{C_3}}{4m^2\ell_{AB}^3\ell \e^2}+\frac{1}{\sqrt{1-u}}\log\(\frac{1+\sqrt{1-u}}{\sqrt u}\)+\gamma\,,\label{eq:re-adjacent-2}
\ee
\esub
and $\ell$ is the total length of the system.
For $m=2$, nice simplifications occur, finding
\be
S^R_{m=2,n}(A:B)=\frac{1}{2}\log \frac{\ell_A\ell_B\ell_{AC_1}\ell_{BC_3}}{\ell_{AB}\ell\e^2}+\gamma_n\,. \label{eq:renyi-re-adjacent-m2}
\ee

\medskip
\noindent \textbf{Taking limits:\,}
In the limit $\ell_{C_1},\ell_{C_3}\rightarrow 0$, which corresponds to the bipartite case, we recover the reflected entropy of pure states \eqref{eq:renyi-re-bipartite}.
Taking limit $\ell_{C_1}, \ell_{C_3}\rightarrow \infty$, the reflected entropies diverge logarithmically due to the zero mode.

For $A$ adjacent to the boundary of the system, that is taking $\ell_{C_1}\rightarrow 0$ ($u\rightarrow 0$), the reflected entropies simplify to
\be
S^R_{m,n}(A:B)
=\frac{1}{2}\log \frac{\ell_A^2\ell_B(m \ell_B \ell+2\ell_{C_3}\ell_A)}  {m \ell\ell_{AB}^2\e^2}+\gamma_n \,. \label{eq:renyi-re-adjacent}
\ee
Curiously, for large systems $\ell\rightarrow \infty$, we get the same result as in the bipartite case $\ell_{C_3}\rightarrow0$.

\smallskip
\subsection{Massive Lifshitz scalar}
\label{sec:massive}

In this section, we study the reflected entropies for the massive deformation of the Lifshitz scalar introduced in \cite{Boudreault:2021pgj}. The calculation of reflected entropies follows exactly as before. The only changes, due to the different propagator \eqref{E:propagator_mass}, are in the matrices encoding adjacency and lengths, whose diagonal elements change from $1/\ell_i$ to $\o \coth \o\ell_i$ and non-diagonal elements from $1/\ell_i$ to $\o\,{\rm csch}\,\o\ell_i$\,. Quite remarkably, for all the cases considered in the previous sections, the reflected entropies for the massive theory are obtained from the massless results by the simple replacement
\ba
\ell_i \rightarrow \sinh\o\ell_i\,,
\ea
and $\e\rightarrow\o\e$. The novel cases come from the ones with periodic BC, which we did not consider for massless Lifshitz scalar due to the presence of a zero mode.

\subsubsection{Disjoint regions on an interval}

The reflected entropies of two disjoint subsystems (see Figure~\scref{fig:disjoint-r}) in the massive case take the same forms as \eqref{eq:renyi-re} and \eqref{eq:re}, with ``effective'' cross-ratio 
\ba
\eta=\frac{\sinh\o\ell_{C_1A}\sinh\o\ell_{BC_3}}{\sinh\o\ell_{C_1AC_2}\sinh\o\ell
_{C_2BC_3}}\,.
\ea
Again, the reflected entropies do not depend on $m$.  In the limit $\o\ell_i\rightarrow 0$  for all $\ell_i$, we recover the massless result.

\medskip
\noindent \textbf{Taking limits:\,}
Consider the configuration in Figure \scref{fig:disjoint-l}.
In the highly gapped regime $\omega \ell_i\rightarrow \infty$ for all $\ell_i$, we have $\eta\sim e^{-2\o \ell_C}$ and the reflected entropies vanish exponentially
\be
\label{eq:disjoint-largemass}
S^R_m(A:B)\simeq \frac  1 2 \o\ell_C e^{-2\o\ell_C}\,, \qquad \o\ell_i\rightarrow\infty\,.
\ee
In the near or far-apart limit, the reflected entropies have the same asymptotics as \eqref{eq:far} and \eqref{eq:near}, with effective cross-ratio $\eta=\frac{\sinh\o\ell_A \sinh\o\ell_B}{\sinh\o\ell_{AC}\sinh\o\ell_{BC}}$\,.

Now consider the general configuration in Figure \scref{fig:disjoint-r}. Since we have a mass gap, there is no divergent zero mode while taking the infinite system limit $\o\ell_{C_1}, \o\ell_{C_3}\rightarrow \infty$. In that case, we have $\eta= e^{-2\o\ell_{C_2}}$ such that the reflected entropies are finite and read
\be
S^R_m(A:B) = \frac{1}{\sqrt{1-e^{-2\o\ell_{C_2}}}}\log\(e^{\o\ell_{C_2}}+\sqrt{e^{2\o\ell_{C_2}}-1}\)-\log\(2\sqrt{e^{2\o\ell_{C_2}}-1}\).
\ee
Interestingly, for infinite systems the reflected entropies do not depend on the lengths $\ell_A,\hspace{1pt}\ell_B$ of the two regions $A, B$, but depend only on the distance $\ell_{C_2}$ between them.
In the highly gapped regime, which coincides with large separation regime, we have \eqref{eq:disjoint-largemass}. In the opposite limit $\o\ell_{C_2}\rightarrow0$, i.e.~small mass or very near $A$ and $B$, we find
\ba
&S^R_m(A:B)\simeq -\frac12\log\o\ell_{C_2}=\frac12\log\frac{\xi}{\ell_{C_2}}   \,,\qquad \o\ell_{C_2}\rightarrow0\,,
\ea
where we have introduced the correlation length $\xi=\o^{-1}$.

\subsubsection{Adjacent regions on an interval}\label{sec:massive-adjacent-interval}

The reflected entropies of two adjacent subsystems (see Figure \scref{fig:adjacent}) in the massive case are obtained from the massless results \eqref{eq:adjacent-t}, \eqref{eq:renyi-re-adjacent-2}, \eqref{eq:re-adjacent-2} by replacing $\ell_i \rightarrow \o^{-1}\sinh\o\ell_i$\,. The massless results are recovered in the limit $\o\ell_i\rightarrow 0$\,.

\medskip
\noindent \textbf{Taking limits:\,}
In the highly gapped regime $\o\ell_i\rightarrow \infty$, writing the correlation length as $\xi=\omega^{-1}$, we have
\be
\label{eq:re-gapped}
S^R_m(A:B)\simeq \log\frac{\xi}{\e}+\gamma-\log2\,.
\ee
The reflected entropy for the gapped theory thus satisfies an area law for adjacent regions. It is also independent of $\ell_A, \ell_B$ in this limit, as expected for massive excitations. For infinite systems, i.e.~$\o\ell_{C_1}, \,\o\ell_{C_3}\rightarrow \infty$, the reflected entropy becomes
\ba
\label{eq:massive-adjacent-asymp}
S^R_{m,n}(A:B) &=\log\frac{\xi}{\e}+ \frac 1 2 \log \frac{(m-2)^2\sinh^2\o\ell_A \sinh^2\o\ell_B}{8m^2e^{\o\ell_{AB}}\sinh^3\o\ell_{AB}}\\
&\qquad\qquad\qquad+\frac{1}{n-1}\log\frac{\big(1+\sqrt{1-u}\big)^{2n}-u^n}{2u^{(n-1)/2}\sqrt{1-u}\big(1+\sqrt{1-u}\big)^n}+\gamma_n\,,\nn
\ea
where
\be
u=\frac{(m-2)^2\sinh\o\ell_A \sinh\o\ell_B}{(2\sinh\o\ell_A+me^{\o\ell_{AB}}\sinh\o\ell_B)(2\sinh\o\ell_B+me^{\o\ell_{AB}}\sinh\o\ell_A )}\,.
\ee

\pagebreak
\subsubsection{Disjoint regions on a circle}

\begin{figure}[t]
\centering
\includegraphics[scale=1.03]{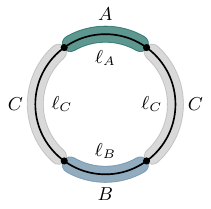}
\caption{Two disjoint regions $A$ and $B$ on finite system with periodic BCs.}
\label{fig:disjoint-circle}
\end{figure}

\begin{figure}[b]
\centering
\includegraphics[scale=1.0]{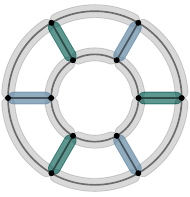}
\caption{Replica graph $\M_{m,3}$ for two disjoint regions on a circle as in Figure \ref{fig:disjoint-circle}. We use a simplified notation here: each gray edge denotes $m/2$ copies of length $\ell_C$ edges connecting the two vertices, while each green (blue) edge denotes $m$ copies of edges of length $\ell_A$ ($\ell_B$).}
\label{fig:disjoint-circle-replica}
\end{figure}

Let us now work on a system with periodic BC, i.e.~a circle. We consider two disjoint regions $A$ and $B$, separated by two intervals of same size $\ell_C$, as can be seen in Figure \ref{fig:disjoint-circle}. We draw the replica graph $\M_{m,3}$ in Figure \ref{fig:disjoint-circle-replica}, which is straightforwardly generalized to arbitrary $n$.
The corresponding partition function reads
\be
\Z_{m,n}=\pi^{2n(1-m)}\(\ell_{A}\ell_B\ell_C^2\)^{-mn/2}\( \det M_{m,n}\)^{-1/2},
\ee
where $M_{m,n}$ is a $4n\times 4n$ matrix of the form
\be
M_{m,n}=
\begin{pmatrix}
X & Y\\
Y & X
\end{pmatrix}.
\ee
Here $X$ is the ``coefficient matrix'' of the vertices on the inner and outer circle of the replica graph $\M_{m,n}$,
while $Y$ is a diagonal matrix connecting the two circles.
We have that $\det M_{m,n}=\det (X+Y)\det(X-Y)$, and both $X+Y$ and $X-Y$ are of the form \eqref{eq:m1} up to a total factor. Using \eqref{eq:det1}, we find
\be
\det M_{m,n}=\(\frac{m\o/2}{\sinh\o\ell_C}\)^{4n}16\sinh^2\hspace{-1.5pt}\(n\,\mathrm{arccosh}\big(\eta_{1}^{-1/2}\big)\hspace{-1.5pt}\)\sinh^2\hspace{-1.5pt}\(n\,\mathrm{arccosh}\big(\eta_{2}^{-1/2}\big)\hspace{-1.5pt}\).
\ee
where
\be
\label{eq:tpm}
\eta_{1}=\frac{\cosh\frac {\o\ell_A} 2 \cosh\frac {\o\ell_B} 2}{\cosh\o\big(\ell_C+\frac {\ell_A} 2 \big)\cosh\o\big(\ell_C+\frac {\ell_B} 2 \big)}\,,\quad \eta_{2}=\frac{\sinh\frac {\o\ell_A} 2 \sinh\frac {\o\ell_B} 2}{\sinh\o\big(\ell_C+\frac {\ell_A} 2 \big)\sinh\o\big(\ell_C+\frac {\ell_B} 2 \big)}\,,
\ee
The reflected entropies for two disjoint subsystems on a circle hence follow as
\bsub
\be
\label{eq:renyi-re-circle}
S^R_{m,n}(A:B)=\sum_{i=1,2}\[\frac{1}{n-1}\log\frac{\big(\sqrt{1-\eta_i}+1\big)^{2n}-\eta_i^{n}}{\big(\big(\sqrt{1-\eta_i}+1\big)^{2}-\eta_i\big)^n}\],
\ee
\be
\label{eq:re-circle}
S^R_m(A:B)=\sum_{i=1,2}\[\frac{1}{\sqrt{1-\eta_i}}\log\(\frac{1+\sqrt{1-\eta_i}}{\sqrt \eta_i}\)  -\log \(2\sqrt{\eta_i^{-1}-1}\)\].
\ee
\esub
In the limit of small mass $\o\ell_i\rightarrow 0$ ($\eta_{1}\rightarrow 1$), the reflected entropies diverge. This is due to the zero mode of the massless theory on a circle.

\medskip
\noindent \textbf{Taking limits:\,}
In the highly gapped limit, we have
\ba
\label{eq:disjoint-circle-largemass}
S^R_m(A:B)\simeq \o\ell_C e^{-2\o\ell_C}\,, \qquad\quad \omega \ell_i\rightarrow \infty\,,
\ea
which is twice \eqref{eq:disjoint-largemass}.
We observe a logarithmic divergence at vanishing separations,
\be
\label{eq:near-circle}
S^R_m(A:B)\simeq -\log\o\ell_C +\frac 1 2\log\frac{\sinh\o\ell_A \sinh\o\ell_B}{4\sinh^2\o\ell_{AB}/2} +2(1-\log 2)\,, \qquad\;\; \o\ell_C\rightarrow 0\,,
\ee
and an exponential decay at large separations,
\be
S^R_m(A:B)\simeq \o\ell_C e^{-2\o\ell_C}\big(1+e^{-\o\ell_{AB}}\big)\,, \qquad \quad  \o\ell_C\rightarrow \infty\,.
\ee

\subsubsection{Adjacent regions on a circle}

\begin{figure}[b]
\centering
\includegraphics[scale=1.03]{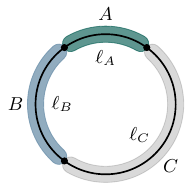}
\caption{Two adjacent regions $A$ and $B$ on finite system with periodic BCs.}
\label{fig:adjacent-circle}
\end{figure}

Finally, let us consider $A$ and $B$ to be adjacent intervals on a circle, as in Figure~\ref{fig:adjacent-circle}.
The replica manifold is similar as that in Figure \ref{fig:adjacent-replica}, only the boundary vertices, which represent Dirichlet BCs, must pair together. Hence, adjacent vertices on the center ``circle'' are connected by $m/2-1$ copies of edge of length $\ell_A+\ell_B$, as well as $m/2$ copies of edge of length $\ell_C$. The corresponding partition function reads
\ba
\Z_{m,n}=\pi^{1-mn}\ell_A^{-n}\ell_B^{-n}\ell_{C}^{-mn/2}(\ell_A+\ell_B)^{-(m-2)n/2}\( \det M_{m,n}\)^{-1/2},
\ea
where $M_{m,n}$ can be found in \eqref{eq:adjacent-circle_Mmn}.
Defining
\be\label{eq:adjacent-u-C}
 u=\frac{\sinh\o\ell_A \sinh\o\ell_B((m-2)\sinh\o\ell_C+m \sinh\o\ell_{AB})^2}{(2\sinh\o\ell_A \sinh\o\ell_C+m \sinh\o\ell_B \sinh\o\ell)(2\sinh\o\ell_B \sinh\o\ell_C+m \sinh\o\ell_A \sinh\o\ell)}\,,
\ee
we obtain the reflected entropies
\bsub
\ba
S^R_{m,n}(A:B)&=-\log \o\e+\frac 1 2 \log \frac{\sinh^2\o\ell_A \sinh^2\o\ell_B\((m-2)\sinh\o\ell_C+m \sinh\o\ell_{AB}\)^2}{16m^2\sinh\o\ell_C \sinh^3\o\ell_{AB} \sinh^2\o\ell/2}\quad\nn\\
&\hspace{1cm}+\frac{1}{n-1}\log\frac{\big(1+\sqrt{1-u}\big)^{2n}-u^n}{2u^{(n-1)/2}\sqrt{1-u}\big(1+\sqrt{1-u}\big)^n}+\gamma_n \,, \label{eq:renyi-adjacent-circle}
\ea
\ba
S^R_m(A:B)&=-\log \o\e+\frac 1 2 \log \frac{\sinh^2\o\ell_A \sinh^2\o\ell_B\((m-2)\sinh\o\ell_C+m \sinh\o\ell_{AB}\)^2}{16m^2\sinh\o\ell_C \sinh^3\o\ell_{AB} \sinh^2\o\ell/2}\quad\nn\\
&\hspace{1cm}+\frac{1}{\sqrt{1-u}}\log\(\frac{1+\sqrt{1-u}}{\sqrt u}\)+\gamma\,. \label{eq:re-adjacent-circle}
\ea
\esub

\medskip
\noindent \textbf{Taking limits:\,}
In the highly gapped regime, we have
\be
S^R_m(A:B)\simeq\log  \frac{\xi}{\e}+\gamma-\log2\,,
\ee
which is identical to \eqref{eq:re-gapped}.
In the bipartite limit $\o\ell_C\rightarrow 0$\,, we find
\be
\label{eq:re-circle-smallc}
S^R_m(A:B)\simeq -\frac 1 2\log\o\ell_C -\log\o\e + \frac 1 2\log\frac{\sinh^2\o\ell_A \sinh^2\o\ell_B}{4\sinh\o\ell \sinh^2\o\ell/2}+1-\log 2+\gamma\,.
\ee
For infinite systems, i.e.~$\o\ell_C\rightarrow \infty$\,, $S^R_{m=1}(A:B)$ has the same asymptotics as \eqref{eq:massive-adjacent-asymp}.

\smallskip
\subsection{Reflected entanglement spectrum}
With the knowledge of all R\'enyi entropies, one can reconstruct the entanglement spectrum. The same applies to the R\'enyi reflected entropies and reflected entanglement spectrum. The latter is defined as the spectrum of the normalized reflected density matrix $\rho_{AA^*}$. Since the reflected entropy of disjoint subsystems is finite, we expect the reflected spectrum to be discrete and cutoff independent (see also \cite{Dutta:2022kge}).

\subsubsection{Entanglement (entropy) spectrum}
Let us start with the entanglement spectrum corresponding to the spectrum of the reduced density matrix $\rho_A$ of some subregion $A$. Consider a region $A$ composed of several intervals, with $k$ boundaries in common with its complement. For Lifshitz theories (both critical and massive) in one spatial dimension, the R\'enyi entropies $S_n(A)$ depend on $n$ through a constant term only \cite{Boudreault:2021pgj}:
\be
S_n(A)=S+\frac{k\log n}{2(n-1)}\,,
\ee
where $S$ is the UV divergent, $n$-independent part. As alluded to above, the entanglement spectrum $P(\l)=\sum_i\delta(\l-\l_i)$ can be related to the moments of the reduced density matrix $R_n\equiv \tr\rho_A^n=\sum_i\l_i^n$, where $\lambda_i$ are the eigenvalues of $\rho_A$, as follows
\be
\label{eq:spec}
R_n= e^{(1-n)S_n}=\frac {e^{(1-n)S}} {n^{k/2}}=\int_0^{\l_{\mathrm{max}}}\l^nP(\l)\,.
\ee
Following \cite{calabrese2008entanglement}, we define
\be
f_j(z)=\frac{1}{\pi}\sum_{j=1}^{\infty}R_j z^{-j}=\frac{1}{\pi}\int d\l \frac{\l P(\l)}{z-\l}\,,
\ee
such that $\l P(\l)=\lim_{\e\rightarrow 0} \im f(\l-i\e)$.
Using $f_j(z)=\pi^{-1}e^S \,\mathrm{Li}_{k/2}(e^{-S}/z)$ and \linebreak\mbox{$\im\, \mathrm{Li}_{k/2} (y+i\e)=\pi\(\ln y\)^{k/2-1}\hspace{-4pt}/\hspace{1pt}\Gamma(k/2)$} for $y\ge 1$ and small positive real $\e$,
we find
\be
P(\l)=\frac{e^{S}\(\ln\l_{\mathrm{max}}/\l\)^{k/2-1}}{\l \Gamma(k/2)}\,,\qquad \l_{\mathrm{max}}=e^{-S}\,.
\ee
The entanglement spectrum is thus continuous, and UV-dependent (through $S$).

\subsubsection{Reflected entanglement spectrum}

\begin{figure}[b]
\centering
\includegraphics[width=\textwidth]{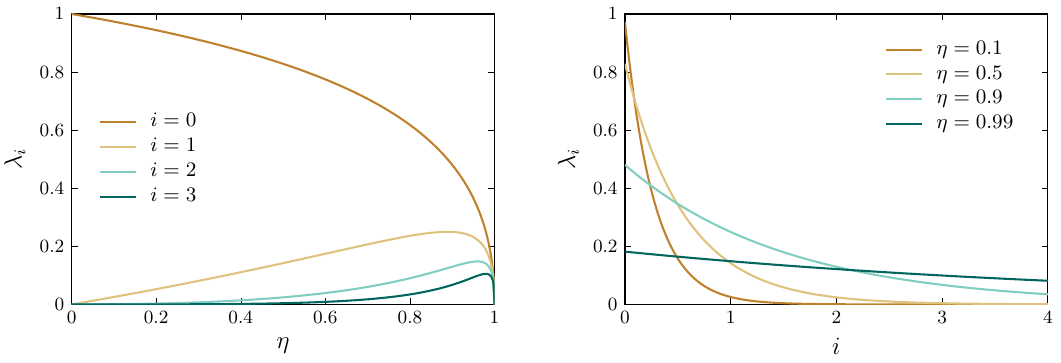}
\caption{Reflected spectrum for two disjoint regions as depicted in Figure \scref{fig:disjoint-l}. \mbox{\textit{Left:}~the largest four eigenvalues \hspace{-.5pt}$\lambda_i$ in the reflected spectrum as functions of cross-ratio $\eta$.} \textit{Right:} the eigenvalues $\lambda_i$ as functions of $i$, for fixed $\eta$. Note that $i$ only takes value in non-negative integers. The spectrum becomes flatter as $\eta\rightarrow 1$.}
\label{fig-spec}
\end{figure}

We now compute the reflected entanglement spectrum of massless Lifshitz scalar field for disjoint regions $A,B$ on an interval (see Figure \scref{fig:disjoint-l}). Using \eqref{eq:renyi-re}, we have
\be
\sum_i\l_i^n=e^{(1-n)S^R_{m,n}}=\frac{(1-q)^n}{1-q^{n}}=\sum_{i=0}^{\infty}((1-q)q^{i})^n\,,
\ee
where we defined 
$q=\eta\(1+\sqrt{1-\eta}\)^{-2}$ with  $\eta=\frac{\ell_A\ell_B}{\ell_{AC}\ell_{BC}}$. Hence we obtain the reflected entanglement spectrum as
\be
\label{eq:spec-massless}
P(\l)=\sum_{i=0}^{\infty} \delta\(\l-(1-q)q^{i}\).
\ee
It is discrete, as expected. We show the first few eigenvalues in Figure \ref{fig-spec}. 

\pagebreak
The reflected entropy is the thermal entropy of a quantum harmonic oscillator with energies $E_l=l+\frac 1 2$ at inverse temperature $\beta=-\log q=2\log \(1+\sqrt{1-\eta}\)-\log\eta$\,:
\be
S^R_m=(1-\beta\partial_\beta)\log Z(\beta)\,,\qquad Z(\beta)=\sum_{l=0}^\infty e^{-\beta (l+\frac 1 2)}=\frac {e^{-\beta/2}}{1-e^{-\beta}}\,.
\ee
The same conclusions hold for the massive Lifshitz scalar, with $\eta=\frac{\sinh\o\ell_A \sinh\o\ell_B}{\sinh\o\ell_{AC}\sinh\o\ell_{BC}}$.

For massive Lifshitz scalar and two disjoint regions on a circle as in Figure \ref{fig:disjoint-circle}, using \eqref{eq:renyi-re-circle} we obtain the reflected entanglement spectrum
\be
P(\l)=\sum_{i,j=0}^{\infty} \delta\(\l-(1-q_1)(1-q_2)q_1^{i} q_2^{j}\),
\ee
where $q_a=\eta_a\(1+\sqrt{1-\eta_a}\)^{-2}$ with $\eta_a$ defined in \eqref{eq:tpm}. The spectrum is, again, discrete.

\smallskip
\subsection{Violation of monotonicity}

\begin{figure}[b]
\centering
\includegraphics[width=\textwidth]{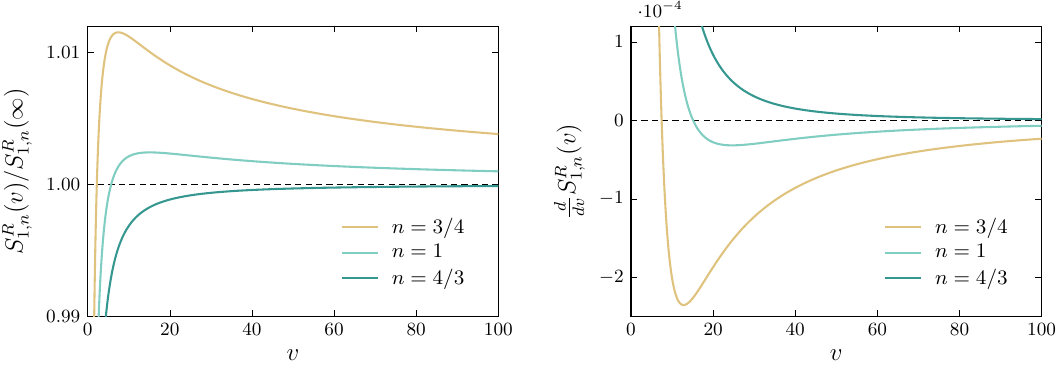}
\caption{Reflected entropies $m=1$ and $n=3/4,\,1,\,4/3$, for two adjacent intervals as in Figure~\ref{fig:adjacent}. \textit{Left:} normalized reflected entropies $S_{1,n}^R(v)/S_{1,n}^R(\scalebox{0.9}{$\infty$})$ as a function of $v=\ell_B/\ell_A$. \textit{Right:} derivative of reflected entropies with respect to $v$, $\frac{d}{dv}S_{m=1,n}^R$, which can be negative for $0<n<4/3$.}
\label{fig-mono}
\end{figure}

A fundamental property a correlation measure $\mathcal{C}$ should satisfy is that it must not increase under the local discarding of information, i.e.~the following inequality must holds
\be\label{monotone_ineq}
\mathcal{C}(A:B\cup D) \geqslant \mathcal{C}(A:B)\,.
\ee
A quantity satisfying this inequality is said to be monotonically nonincreasing under partial trace, or simply `monotonic'.

It has recently been shown that the reflected entropy is not a measure of physical correlations since it is not monotonic \cite{Hayden:2023yij}. However, for holographic theories and free fields (scalars and Dirac fermions) \cite{Bueno:2020fle}, the reflected entropy still seems to satisfy monotonicity. 
For Lifshitz theories, using our exact results for two adjacent intervals \eqref{eq:renyi-re-adjacent-2} and \eqref{eq:re-adjacent-2}, we show in Figure \ref{fig-mono} that monotonicity is violated. This provides a counterexample to monotonicity of reflected entropy for a free theory.

\pagebreak
For the critical theory, it can be rigorously shown that monotonicity is violated for R\'enyi indices $0< n<4/3$. Consider the large system limit $\ell_{C_1},\ell_{C_3}\rightarrow\infty$ of \eqref{eq:renyi-re-adjacent-2} and \eqref{eq:re-adjacent-2}, and let $v=\ell_B/\ell_A$. The derivative of the reflected entropy with respect to $v$ is found to be
\ba
\lim_{v\to\infty}\frac{d}{dv}S_{m=1,n}^R=
\begin{cases}
\displaystyle-\frac{8^{-n}n}{1-n}\hspace{1pt}v^{-(n+1)} + \cdots\,, \qquad &  n<1\,,\vspace{7pt}\\
\displaystyle-\frac{1}{8}\hspace{1pt}v^{-2}\log v + \cdots\,, & n=1\,,\vspace{7pt}\\
\displaystyle-\frac{4-3n}{8(n-1)}\hspace{1pt}v^{-2} + \cdots\,, & n>1\,,
\end{cases}
\ea
which is negative for large $v$ in the range $0<n<4/3$. It can be shown to be positive for $n\geqslant 4/3$ for any $v>0$. This implies that the reflected entropy is not monotonic.

\section{Comparison with mutual information: the Markov gap}
\label{sec:mi}

In this section, we study a tripartite entanglement measure recently proposed \cite{Hayden:2021gno,Akers:2019gcv,Zou:2020bly}, the Markov gap $M$, defined as the difference of the reflected entropy with mutual information~$I$
\ba\label{Mgap}
M(A : B) \equiv S^R(A : B) - I(A : B)\,.
\ea
The reflected entropy is lower bounded by mutual information \cite{Dutta:2019gen}, hence the Markov gap is a nonnegative quantity $M(A : B) \ge 0$.
As shown in \cite{Zou:2020bly}, a nonvanishing $M$ for contiguous subsystems $A, B, C$ signals irreducible tripartite entanglement, beyond bipartite entanglement and GHZ-type entanglement.

Here, we show that for Lifshitz groundstates, the Markov gap is strictly positive $M>0$, both for critical and gapped states (except for vanishing correlation length where it is zero), implying nontrivial tripartite entanglement. It is interesting that, despite the fact that two disjoint regions are not entangled \cite{Boudreault:2021pgj}, a form of nonbipartite entanglement is still shared between the three complementary subsystems.
Before proceeding, let us recall certain results about mutual information in Lifshitz groudstates.

\subsection{Mutual information}
For the sake of being self-contained, we give here the (R\'enyi) mutual information for the various cases considered in this paper, most of which can be found in \cite{Boudreault:2021pgj}. The R\'enyi mutual information $I_n(A:B)$ between two subsystems $A$ and $B$  is defined as
\begin{align}
I_n(A:B) = S_n(A) + S_n(B) - S_n(A\cup B)\,, \quad
\end{align}
where $S_n(X)$ is the R\'enyi entropy for the subsystem $X$.
Mutual information is obtained in terms of entanglement entropies by taking the $n\rightarrow 1$ limit, $I(A:B) = \lim_{n\rightarrow 1} I_n(A:B)$,
and is a measure of total correlations between $A$ and $B$.

The (R\'enyi) mutual information of two disjoint intervals in the bulk of a finite system with Dirichlet BCs (see Figure~\ref{fig:disjoint-r}) reads
\ba
\label{eq:MI-disj-D}
I_n(A:B) = \frac12 \log\frac{\sinh\o\ell_{C_1AC_2}\sinh\o\ell_{C_2BC_3}}{\sinh\o\ell_{C_2}\sinh\o\ell}=\frac 1 2\log \frac{1}{1-\eta}\,,
\ea
with $\eta=\frac{\sinh\o\ell_{C_1A}\sinh\o\ell_{BC_3}}{\sinh\o\ell_{C_1AC_2}\sinh\o\ell
_{C_2BC_3}}$, while for two adjacent subsystems it is given by 
\ba
\label{eq:MI-adj-D}
I_n(A:B) = \frac12 \log\frac{\sinh\o\ell_{C_1A}\sinh\o\ell_{BC_3}\sinh\o\ell_A\sinh\o\ell_B}{(\o\e)^2\sinh\o\ell\sinh\o\ell_{AB}}+\gamma_n\,.
\ea

On the circle with periodic BC, the R\'enyi mutual information of two disjoint regions (as in Figure \ref{fig:disjoint-circle}) reads
\ba\label{eq:MI-disj-P}
I_n(A:B)=\frac 1 2\log\frac{\sinh\o\ell_{ACC}\sinh\o\ell_{BCC}}{4(\sinh\o\ell_C\sinh\o\ell/2)^2}=\sum_{i=1,2}\frac 1 2\log \frac{1}{1-\eta_i}\,,
\ea
with $\eta_{1,2}$ defined in \eqref{eq:tpm}, while for two adjacent subsystems it is given by
\ba
\label{eq:MI-adj-P}
I_n(A:B) = \frac12 \log\frac{\sinh\o\ell_A\sinh\o\ell_B\sinh\o\ell_{AC}\sinh\o\ell_{BC}}{\sinh\o\ell_{AB}\sinh\o\ell_C\sinh^2\o\ell/2}-\log 2\o\e+\gamma_n\,.
\ea
In all cases, mutual information for the critical Lifshitz scalar is obtained by setting $\o\rightarrow0$.

We note that, interestingly, the $m=2$ reflected entropies equal mutual information:
\be
\label{eq:S_2-MI}
\begin{aligned}
S^R_{m=2,n}(A:B)= I_n(A:B)\,, \qquad\quad &A,B\;\, \text{adjacent on an interval}\,,\\
\qquad S^R_{m=2,n=2}(A:B)= I_{n=2}(A:B)\,, \qquad\quad &A,B\;\, \text{adjacent on a circle}\,.
\end{aligned}
\ee

\smallskip
\subsection{The Markov gap}
We are now in the position to compute the Markov gap, as defined in \eqref{Mgap}, using the results of Section~\ref{sec:11} on reflected entropy and mutual information given just above.

\subsubsection{Disjoint regions}

For two disjoint regions in the bulk of a finite system with Dirichlet BCs, as in Figure~\ref{fig:disjoint-r}, we find
\ba
M(A:B) = \frac{1}{\sqrt{1-\eta}}\log\(\frac{1+\sqrt{1-\eta}}{\sqrt\eta}\)  +\log\sqrt\eta -\log2\,.
\ea
In this case, $M(A:B)$ is a positive-definite, monotonically increasing function of $\eta$. For infinite systems $\ell_{C_1},\ell_{C_3}\rightarrow\infty$, it depends only on the separation $\ell_{C_2}$ since $\eta=e^{-2\o\ell_{C_2}}$. 

The Markov gap reaches its maximum value when the distance between the two regions goes to zero, i.e.~$\ell_{C_2}\rightarrow0$ or $\eta\rightarrow1$
\ba
\qquad M(A:B)=1-\log2\,, \qquad \eta\rightarrow1\,.
\ea
In the opposite $\eta\rightarrow 0$ limit, which corresponds to either large separation or large mass, the Markov gap vanishes as
\ba\label{MG-disj-D}
\qquad M(A:B)\simeq -\frac14 \eta\log\eta\,, \qquad \eta\rightarrow0\,.
\ea
Note that in this $\eta\rightarrow 0$ limit, we have
$I(A:B)\simeq \frac{1}{2}\eta$, and $S^R_m(A:B)\simeq -\frac14 \eta \log \eta$, which matches the conjecture of \cite{Bueno:2020fle} (see also \cite{Camargo:2021aiq}) for CFTs that
\be\label{eq:remi}
S^R(A:B)\simeq -\frac12 I(A:B) \log \eta \,, \qquad \eta\rightarrow0\,.
\ee
It is interesting that for Lifshitz theory, although not a CFT, this relation is also satisfied.

\smallskip
For two disjoint regions on a circle, see Figure~\ref{fig:disjoint-circle}, the Markov gap is given by
\ba
M(A:B) = \sum_{i=1,2} \frac{1}{\sqrt{1-\eta_i}}\log\(\frac{1+\sqrt{1-\eta_i}}{\sqrt\eta_i}\)  +\log\sqrt\eta_i -\log2\,,
\ea
with $\eta_{1,2}$ defined in \eqref{eq:tpm}.
In the critical massless case, i.e.~$\eta_1=1$ and $\eta_2=\frac{\ell_A\ell_B}{(\ell_A+2\ell_C)(\ell_B+2\ell_C)}$, there is no divergence due to the zero mode as for the reflected entropy and mutual information. Furthermore, at large separation $\ell_C\rightarrow\infty$ the Markov gap does not vanish, but goes to $M(A:B)=1-\log 2$. Note that the massless limit and large separation limit do not commute. 
In the small separation $\ell_C\rightarrow 0$, the Markov gap also goes to a constant $M(A:B)= 2(1-\log 2)$. This suggests that for two disjoint regions with $k$ pairs of very near boundaries, the gap is $k(1-\log 2)$. Finally, at large separation $\o\ell_C\rightarrow \infty$ or large mass, we find twice \eqref{MG-disj-D} and \eqref{eq:remi} for the asymptotics of $M$ and $S^R$, respectively.

\subsubsection{Adjacent regions}

Let us now turn to adjacent subsystems. For two adjacent regions in the bulk of a finite system with Dirichlet BCs, as depicted in Figure~\ref{fig:adjacent}, we obtain
\ba\label{MG_1d_m}
M(A:B) =\frac12\log\frac{\sinh\o\ell_A\sinh\o\ell_B\sinh\o\ell_{C_1}\sinh\o\ell_{C_3}}{4\sinh^2\o\ell_{AB}\sinh\o\ell_{C_1A}\sinh\o\ell_{BC_3}}+ \frac{1}{\sqrt{1-u}}\log\(\frac{1+\sqrt{1-u}}{\sqrt u}\),
\ea
where $u$ is given in \eqref{eq:adjacent-t} with the replacement $\ell_i \rightarrow \sinh\o\ell_i$ and $m=1$.
In the critical massless case, for infinite systems $\ell_{C_1}, \ell_{C_3} \rightarrow \infty$, the Markov gap is a positive UV-finite function of the ratio $v\equiv\ell_B/\ell_A$, symmetric in $v\rightarrow1/v$, i.e.
\ba
M(v)=\frac{1}{2} \log\(\frac{v}{4(1+v)^2}\)+\sqrt{\frac{(2+v)(1+2v)}{2(1+v)^2}}\, \mathrm{arcsinh}\sqrt{2v^{-1}(1+v)^2}\,, 
\ea
and takes value between $\frac12\log2$ for $v\rightarrow0,\infty$ and $\frac{3}{2 \sqrt{2}} \log \left(3+2\sqrt2\right)-2\log 2$ for $v=1$. It is interesting to compare this result for Lifshitz theories with the CFT one, which is universal \cite{Zou:2020bly}: $M(A:B) = (c/3) \log2$ (with $c$ the central charge). For CFTs, the Markov gap of two adjacent subsystems is independent of the relative size of those in the scaling limit, contrarily to Lifshitz theory.
Moreover, a nonvanishing $M$ indicates irreducible tripartite entanglement between the three subsystems $A, B, C$, beyond GHZ-like entanglement.

In the highly gapped case where the size of each subsystem is much larger than the correlation length $\xi=\o^{-1}$, the Markov gap vanishes exponentially
\ba
\qquad M(A:B) \simeq e^{-2\o\ell_A}+ e^{-2\o\ell_B}\,,\qquad \o\ell_i\rightarrow\infty\,,
\ea
which is expected for gapped systems with vanishing correlation length \cite{Zou:2020bly}.

\smallskip
On a circle with periodic BC (Figure~\ref{fig:adjacent-circle}), the Markov gap is given by 
\ba
M(A:B) =\frac12\log\frac{\sinh\o\ell_A\sinh\o\ell_B(\sinh\o\ell_{C}-\sinh\o\ell_{AB})^2}{4\sinh^2\o\ell_{AB}\sinh\o\ell_{AC}\sinh\o\ell_{BC}}+ \frac{1}{\sqrt{1-u}}\log\(\frac{1+\sqrt{1-u}}{\sqrt u}\)\hspace{-2pt},
\ea
with $u$ defined in \eqref{eq:adjacent-u-C} setting $m=1$.
In the two limits \mbox{$\ell_C\rightarrow\infty$} and $\o\rightarrow\infty$, it displays the same asymptotics as that for Dirichlet BCs.

\section{Computable cross-norm negativity and operator entanglement entropy}
\label{sec:oe}

The reflected entropies are related to two other information-theoretic quantities (see, e.g.~\cite{Berthiere:2023gkx}), namely the computable cross-norm or realignment (CCNR) negativity and the operator entanglement entropy (OEE). As a byproduct of our results on reflected entropy, we provide exact formulas for the CCNR negativity and OEE for Lifshitz theories.

\subsection{Computable cross-norm negativity}

The CCNR negativity is based on the CCNR separability criterion \cite{Rudolph2002,2002quant.ph..5017C}. It has recently been discussed in a variety of contexts \cite{Yin:2022toc,Milekhin:2022zsy,Yin:2023jad,Berthiere:2023gkx}, and it is expressed in terms of the realignment matrix $R$ of a bipartite density matrix $\rho_{AB}$, i.e.~$\bra a\bra{a'}R\ket b\ket{b'}=\bra a\bra{b}\rho_{AB}\ket {a'}\ket{b'}$ where $\ket a\hspace{-2pt}, \ket b$ are basis of $A,B$. The R\'enyi CCNR negativities are defined as $\mathcal{E}_n(A:B)=\log\tr(RR^\dagger)^n$, and the limit $n\to1/2$ yields the CCNR negativity $\mathcal{E}$. For separable states, $\E\le0$. 

A generalization of CCNR negativity has been introduced in \cite{Berthiere:2023gkx} using the realignment matrix of $\rho_{AB}^{m/2}$. The $(m,n)$--Rényi CCNR negativity can then alternatively be expressed in terms of the $(m,n)$--Rényi reflected entropy as \cite{Berthiere:2023gkx}
\be\label{CCNR_RE}
\mathcal{E}_{m,n}(A:B) = (1-n)S^R_{m,n}(A:B) + n(1-m)S_m(A\cup B) \,,
\ee
such that setting $m=2$ we get the Rényi CCNR negativity $\E_n$. The $(m,n)$--Rényi CCNR negativity can be viewed as the unnormalized $(m,n)$--Rényi reflected entropy.

\medskip
\noindent \textbf{Disjoint regions.\,} %
Using our results on reflected entropy, we obtain the CCNR negativities for disjoint regions on a interval (Figure \scref{fig:disjoint-r}) as
\bsub
\be
\E_{m,n}(A:B)=-\log\frac{\big(\sqrt{1-\eta}+1\big)^{2n}-\eta^{n}}{\big(\big(\sqrt{1-\eta}+1\big)^{2}-\eta\big)^n}+n(1-m)\hspace{-2pt}\[\frac 1 2 \log\frac{\ell_A\ell_{B}\ell_C}{\ell\e^2}+ \gamma_m\]\hspace{-2pt},
\ee
\be
\E(A:B)=-\frac 1 2 \log \big(\eta^{-1/2}-1\big) -\frac 1 2 \log\frac{2\pi\ell_A\ell_B}{\ell\e}\,.
\ee
\esub
where $\eta=\frac{\ell_A\ell_B}{\ell_{AC}\ell_{BC}}$.
The CCNR negativity $\E$ is negative, as long as the cutoff is much smaller than all other length scales. This is consistent with the separability of the density matrix $\rho_{AB}$ for disjoint subsystems \cite{Boudreault:2021pgj}.

When adding mass, the CCNR negativity becomes
\ba
\E(A:B)=-\frac 1 2 \log \big(\eta^{-1/2}-1\big)  -\frac 1 2 \log\frac{2\pi \sinh\o\ell_A \sinh\o\ell_B}{\o\e \,\sinh\o\ell}\,,
\ea
with $\eta=\frac{\sinh\o\ell_A \sinh\o\ell_B}{\sinh\o\ell_{AC}\sinh\o\ell_{BC}}$. It has qualitatively the same behavior as the massless case.

\medskip
\noindent \textbf{Adjacent regions.\,} %
For adjacent regions (Figure \scref{fig:adjacent}), we find
\be
\E_{m,n}(A:B)=(1-n)\[\frac{1}{2}\log \frac{\ell_A\ell_B\ell_{AC_1}\ell_{BC_3}}{\ell_{AB}\ell\e^2}+\gamma_n\]+n(1-m)\[\frac 1 2 \log\frac{\ell_{AB}\ell_{C_1}\ell_{C_3}}{\ell\e^2}+\gamma_m\]\hspace{-2pt},
\ee
\be\label{CCNR-adj-finite}
\E(A:B)=\frac 1 4\log \frac{4\ell_A\ell_B\ell_{AC_1}\ell_{BC_3}}{\ell_{C_1}\ell_{C_3}\ell^2_{AB}}\,.
\ee
For infinite systems, i.e.~$\ell_{C_1},\ell_{C_3}\rightarrow\infty$, the CCNR negativity simplifies as
\ba\label{CCNR-adj}
\qquad\qquad\E(A:B)=\frac 1 4\log \frac{4\ell_A\ell_B}{(\ell_A+\ell_B)^2}\,, \qquad\qquad \ell_{C_1},\ell_{C_3}\rightarrow\infty\,.
\ea
Adding a mass we have
\ba\label{CCNR-adj-finite-m}
\qquad\quad\E(A:B)&=\frac 1 4\log \frac{4\sinh\o\ell_A \sinh\o\ell_B \sinh\o\ell_{AC_1} \sinh\o\ell_{BC_3}}{\sinh\o\ell_{C_1}\sinh\o\ell_{C_3}\sinh^2\o\ell_{AB}}\,,\\
&=\frac 1 4\log \frac{4\sinh\o\ell_A \sinh\o\ell_B}{\sinh^2\o\ell_{AB}}+\frac14\o(\ell_A+\ell_B)\,, \qquad\quad \ell_{C_1},\ell_{C_3}\rightarrow\infty\,.\label{CCNR-adj-m}
\ea
Quite surprisingly, the CCNR negativity for two adjacent regions is cutoff independent. In fact in this case, the $(m,n)$--Rényi CCNR negativity is UV-finite for $mn=1$. For finite systems, expressions \eqref{CCNR-adj-finite} and \eqref{CCNR-adj-finite-m} can be negative or positive. However, for infinite systems in the critical massless case, the CCNR negativity \eqref{CCNR-adj} is nonpositive and thus fails to detect entanglement. Finally, CCNR negativity does not satisfy monotonicity as is evident from \eqref{CCNR-adj}. The nonmonotonicity of CCNR negativity was proven in \cite{Berthiere:2023gkx}.

\subsection{Operator entanglement entropy}

The Rényi OEE \cite{Zanardi:2001zza,2007PhRvA..76c2316P,Dubail:2016xht,Rath:2022qif}, which we denote as $E_n$, is the (Rényi) Shannon entropy of the squared probability distribution values of the operator Schmidt decomposition of a (normalized) bipartite density matrix $\rho_{AB}$. Hence, the OEE is the reflected entropy of $\rho_{AB}$, that is 
\be
E_n(A:B)=S^R_{m=2,n}(A:B)\,.
\ee
Our results for the $m=2$ reflected entropy thus straightforwardly apply to the OEE. In particular, we have shown that for disjoint intervals, the reflected entropies for Lifshitz theories are independent of $m$, see \eqref{eq:renyi-re} and \eqref{eq:renyi-re-circle}. 
For two adjacent intervals on a finite system with Dirichlet BCs, the $m=2$ reflected entropy exactly matches mutual information, thus so does the OEE.

\section{Towards $2+1$ dimensions}
\label{sec:21}

In this section, we compute the reflected entropies for $2+1D$ Lifshitz groundstates on a cylinder.
The idea is that, given the symmetry of the system in two spatial dimensions, one can perform a dimensional reduction along the compact direction to get a tower of $1d$ free massive theories. This allows us to compute $2d$ entropies by summing up (massive) $1d$ entropies.
We first reproduce the $2d$ entanglement entropies from the $1d$ results, and then move on to reflected entropies in $2d$ in certain limits.

\subsection{Entanglement entropy on a cylinder}

Consider the $(2+1)$-dimensional non-compact free real Lifshitz scalar with Hamiltonian
\be
H_{2d}=\frac 1 2 \int d x dy \(\Pi^2+(\nabla^2 \phi)^2 \)
\ee
on a cylinder of total length $\ell=\ell_A+\ell_B$ with Dirichlet BC at both ends, which corresponds to Figure \scref{fig:bipartite} times a circle $S^1$ with length $\ell_y$. The entanglement entropy of region $A$ can be calculated using the replica trick and field redefinitions \cite{Fradkin:2006mb}.  The universal (constant) term in the entanglement entropy is \cite{Zhou:2016ykv}
\be
\label{eq:ee}
S^{2d}(A)= \log \frac{\eta(2w\tau)\eta(2\tau-2w\tau)}{\eta(2\tau)}+\frac 1 2 \log\( 2w(1-w)|\tau|\),
\ee
where $w=\ell_A/\ell$ and $\tau=i\ell/\ell_y$. 
Let us now reproduce this $2d$ result using that of $1d$.

The partition function on the replica manifold for the $2d$ theory reads
\be
\Z^{2d}_{n}=\int_{\M_n} D\phi \, e^{-S_{\rm cl}[\phi]}\,, \qquad\quad S_{\rm cl}[\phi]=\int dx dy (\nabla \phi)^2\,.
\ee
On a cylinder, any replica manifold $\M_n$ is a graph times $S^1$. This $S^1$ symmetry allows us to perform a dimensional reduction (in the transverse $y$-direction) and to write
\be
\label{eq:decomp}
S_{\rm cl}[\phi]=\ell_y\int dx (\p_x\phi_0)^2+2\ell_y\sum_{k=1}^{\infty}\int dx\( |\p_x\phi_k|^2+\o_k^2|\phi_k|^2\)\hspace{-1pt}.
\ee
We have expanded $\phi(x,y)=\phi_0(x)+\sum_{k\neq 0} e^{iky}\phi_k(x)$, and the frequencies are $\o_k=\frac{2\pi k}{\ell_y}$.
Each mode $k$ is a complex scalar field, which can be viewed as two real scalars.
Hence
\be
\label{eq:2dz}
\Z^{2d}_n=\Z_{0,n}\( \prod_{k=1}^{\infty} \Z_{k,n}\)^2,
\ee
where $\Z_0$ and $\Z_k$ are quantum mechanical partition function of free particle and harmonic oscillators.
For the entropies we have
\be\label{eq:dim_reduc}
S^{2d}(A)=S_0(A)+2\sum_{k=1}^{\infty} S_k(A)\,,
\ee
where $S_0(A)$ and $S_k(A)$ are the $1d$ entanglement entropies of massless and massive Lifshitz scalar with mass $\o_k$, respectively, which are given by
\be
S_0(A)=\frac 1 2 \log \frac{\ell_A(\ell-\ell_A)}{\ell}\,,\qquad
S_k(A)=\frac 1 2 \log\frac{\sinh \o_k\ell_A\sinh\o_k (\ell-\ell_A)}{\o_k\sinh\o_k \ell}\,.
\ee
Here we have discarded nonuniversal constant terms.
Using $\eta(\tau)=q^{\frac 1 {24}}\prod _{n=1}^{\infty} (1-q^n)$, with $q=e^{2\pi i \tau}$ and the zeta-function regularization, i.e.~$1+2+3+\cdots=-1/12$, we can rewrite the sums of $\log\sinh$-terms as $\eta$-functions. Further, making use of $\prod_{n=1}^{\infty} (a n)=\sqrt{2\pi/a}$, the sum of $\log\o_k$ can be computed in closed form, and we reproduce 
\eqref{eq:ee} exactly.

The entanglement entropy of a bulk cylinder whose dimensional reduction corresponds to region $C$ in Figure \scref{fig:disjoint-l} can be calculated in the same fashion. Using the $1d$ results \cite{Boudreault:2021pgj}
\be
S_0(C)=\frac 1 2 \log \frac{\ell_A\ell_B\ell_C}{\ell}\,,\qquad
S_k(C)=\frac 1 2 \log\frac{\sinh \o_k\ell_A\sinh\o_k \ell_B\sinh\o_k\ell_C}{\o_k^2\sinh\o_k \ell}\,,
\ee
we obtain
\be
\label{eq:ee2}
S^{2d}(C)= \log \frac{\eta(2w_A\tau)\eta(2w_B\tau)\eta(2w_C\tau)}{\eta(2\tau)}+\frac 1 2 \log\( 4w_Aw_Bw_C|\tau|^2\),
\ee
where $\tau=i\ell/\ell_y$ and $w_i=\ell_i/\ell$\,.
\subsection{Reflected entropy on a cylinder}

Let us first consider $A, B$ disjoint on a cylinder, whose dimensional reduction leads to Figure~\scref{fig:disjoint-l}, and the limiting regimes where the reflected entropies get simplified to allow for an analytic treatment.
In the limit
\be
\ell_A\ll\ell_y\,,\ell_C\sim\ell_y\,,\quad \mathrm{or}\quad\ell_B\ll\ell_y\,,\ell_C\sim\ell_y\,, \quad \mathrm{or}\quad \ell_C\gg\ell_y\,,
\ee 
the $1d$ massive reflected entropies vanish. Using \eqref{eq:dim_reduc}, the $2d$  reflected entropy is then entirely given by the $1d$ massless contribution
\be
\qquad\quad S^{R}_m(A:B)=\frac{1}{\sqrt{1-\eta}}\log\(\frac{1+\sqrt{1-\eta}}{\sqrt\eta}\)  -\log \(2\sqrt{\eta^{-1}-1}\), \qquad d=2\,,
\ee
with $\eta=\frac{\ell_A\ell_B}{\ell_{AC}\ell_{BC}}$. In all these three limits, mutual information of the massive theory also vanishes, and the $2d$ mutual information is given by the massless contribution as well. Thus, we expect relation \eqref{eq:remi} to also hold for $2+1D$ Lifshitz theory, at least in the above three limits.

For $A,B$ adjacent on a cylinder, whose dimensional reduction leads to Figure \scref{fig:adjacent}, taking the limits
\be\label{limit_adj}
\ell_A\gg\ell_y \quad \mathrm{and}\quad\ell_B\gg\ell_y\,,
\ee
the reflected entropies for the $1d$ massive theory behaves as $-\log \o_k$ (see \eqref{eq:re-gapped}), whose sum leads to $-(1/2)\log\ell_y$. Note that any constant terms in the $1d$ entropies will drop out after regularization. Using \eqref{eq:dim_reduc}, the (universal term in) the $2d$ reflected entropy reads
\be
\qquad S^{R}_m(A:B)=\frac{1}{\sqrt{1-u}}\log\(\frac{1+\sqrt{1-u}}{\sqrt u}\)+\frac 1 2 \log \frac{(m-2)^2\ell_A^2\ell_B^2\ell_{C_1}\ell_{C_3}}{m^2\ell_{AB}^3\ell \ell_y^2} \,, \qquad d=2\,,
\ee
with $u$ defined in \eqref{eq:adjacent-t}. As in the $1d$ case, the reflected entropy does not satisfy monotonicity in $2d$. Mutual information in this limit can be calculated either using \eqref{eq:ee2} or by summing the $1d$ expressions, finding
\ba
I^{2d}(A:B)&=\frac{1}{2}\log\frac{4\ell_{AC_1}\ell_{BC_3}\ell_A\ell_B}{\ell\ell_{AB}\ell_y^2}\,.
\ea
Only the massless part contributes to the Markov gap in the limit \eqref{limit_adj}:
\ba
M(A:B) &=\frac12\log\frac{\ell_A\ell_B\ell_{C_1}\ell_{C_3}}{4\ell_{AB}^2\ell_{C_1A}\ell_{BC_3}}+ \frac{1}{\sqrt{1-u}}\log\(\frac{1+\sqrt{1-u}}{\sqrt u}\),
\ea
which indeed corresponds to \eqref{MG_1d_m} in the $1d$ massless case. It is positive, indicating nontrivial tripartite entanglement.

More generally, for adjacent regions on a cylinder, the reflected entropies for $m=2$ simplify drastically, see \eqref{eq:renyi-re-adjacent-m2}, and are equal to mutual information exactly (see \eqref{eq:S_2-MI}). The $2d$ reflected entropies for $m=2$ then follow straightforwardly, as well as the $2d$ operator entanglement entropies since $E_n(A:B)=S^{R}_{m=2,n}(A:B)=I_n(A:B)$.

\section{Discussion}\label{conclu}

In this paper, we have investigated the reflected entropy, and its R\'enyi generalization, for Lifshitz groundstates. We focused on the critical free $z=2$ Lifshitz scalar theory in $1+1$ dimensions, as well as its massive deformation, and considered their groundstates defined on tripartite systems. We computed the reflected entropies for two intervals, either adjacent or disjoint, and obtained explicit formulas for general R\'enyi indices $m,n$. In the case of two disjoint intervals, we found that the reflected entropies are UV-finite and do not depend on the R\'enyi index $m$. We have also computed the corresponding reflected entanglement spectrum, finding a discrete set of eigenvalues which is that of a thermal density matrix. For two adjacent regions, the reflected entropy is not monotonic under partial trace, providing the first counterexample to  monotonicity in free theories.

Although not a genuine bipartite correlation measure since it is not monotonic, the reflected entropy has a meaningful relationship to multipartite entanglement. Indeed, the Markov gap, defined as the difference between the reflected entropy and mutual information, can detect beyond bipartite and GHZ-type entanglements. We studied this Markov gap and found it to be positive, signaling irreducible tripartite entanglement in Lifshitz groundstates. 
Despite the fact that two disjoint regions are not entangled \cite{Boudreault:2021pgj}, a form of nonbipartite entanglement thus subsists among the three complementary subsystems.

As a byproduct of our results on reflected entropy, we provided exact formulas for two other entanglement-related quantities, namely the CCNR negativity and the operator entanglement entropy. Quite surprisingly, the CCNR negativity is shown to be UV-divergent for two disjoint subsystems, but UV-finite for two adjacent intervals. It is also found to be nonmonotonic under partial trace.

Finally, via dimensional reduction, we obtained analytic expressions for the
reflected entropy in certain limits in $2+1$ dimensions, using our results for $(1+1)$--dimensional massive Lifshitz scalar field.

There are several future avenues worth exploring. An important direction would be to investigate further the reflected entropy in higher-dimensional Lifshitz theories. We took here a decisive step in that direction, though much remains to be done. For example, the study of skeletal regions \cite{Berthiere:2021nkv}, i.e.~subregions $A,B$ that have no volume, could lead to new analytical results in two or more spatial dimensions.
Reflected entropy bears importance in the holographic context, where it is dual to the area of entanglement wedge cross-section. It would be interesting to relate our results to holography, see, e.g., \cite{BabaeiVelni:2019pkw,BabaeiVelni:2023cge}.
Finally, it is worth studying reflected entropy for the more general class of Rokhsar-Kivelson states--to which belong Lifshitz groundstates--to further explore the nature of tripartite entanglement in such states (see \cite{Boudreault:2021pgj,Parez:2022ind} for recent developments on their entanglement structure).

\begin{acknowledgments}
It is a pleasure to thank Gilles Parez for fruitful discussions and collaboration on related work. C.B.~was supported by a CRM-Simons Postdoctoral Fellowship at the Universit\'e de Montr\'eal. B.C.~and H.C.~are supported by NSFC Grant No.~11735001.
\end{acknowledgments}

\bibliographystyle{JHEP}
\bibliography{ref}

\newpage
\appendix
\section{Matrices and determinants}
\label{app:det}
In this appendix, we list the matrices and their determinants used in the main text.

\subsection{Matrices of replica graphs}

In our calculations, partition functions $\Z_{m,n}$ corresponding to replica graphs of geometric configurations are given in terms of the determinants of certain matrices $M_{m,n}$ as
\be
\Z_{m,n}\propto\( \det M_{m,n}\)^{-1/2}.
\ee
We list here such matrices for the different cases we consider.
\medskip

\noindent\textbf{Disjoint regions on an interval.\,} The geometry and replica graph $\M_{m,n}$ are drawn in Figure~\ref{fig:disjoint-replica}, and the corresponding matrix $M_{m,n}$ is a $2n\times 2n$ matrix of the form
\be
M_{m,n}=
\begin{pmatrix}
\frac{m}{\ell_A}+\frac{m}{\ell_C} & -\frac{m/2}{\ell_C} & 0&\cdots&0&-\frac{m/2}{\ell_C}\\
-\frac{m/2}{\ell_C}&\frac{m}{\ell_B}+\frac{m}{\ell_C} & -\frac{m/2}{\ell_C}&0&\cdots&0\\
0&-\frac{m/2}{\ell_C}&\frac{m}{\ell_A}+\frac{m}{\ell_C} & -\frac{m/2}{\ell_C}&\ddots&\vdots\\
\vdots&\ddots&\ddots&\ddots&\ddots&0\\
0&\cdots&0&-\frac{m/2}{\ell_C}&\frac{m}{\ell_A}+\frac{m}{\ell_C} & -\frac{m/2}{\ell_C}\\
-\frac{m/2}{\ell_C}&0&\cdots&0&-\frac{m/2}{\ell_C}&\frac{m}{\ell_B}+\frac{m}{\ell_C} 
\end{pmatrix}.
\label{eq:disjoint-matrix}
\ee

\smallskip

\noindent\textbf{Adjacent regions on an interval.\,} The geometry and replica graph $\M_{m,3}$ (which can easily be generalized to arbitrary $n$) are drawn in Figures~\ref{fig:adjacent} and \ref{fig:adjacent-replica}, and the corresponding $M_{m,n}$ is the $(2n+2)\times (2n+2)$ matrix:
\ba
\scalebox{0.8}{$
\begin{pmatrix}
\frac{n}{\ell_A}+\frac{n}{\ell_B}& -\frac{1}{\ell_A} &-\frac{1}{\ell_B}& -\frac{1}{\ell_A} &-\frac{1}{\ell_B}&\cdots&-\frac{1}{\ell_B}&0\vspace{2pt}\\
-\frac{1}{\ell_A}&\frac{m}{\ell_{C_1}} +\frac{2}{\ell_{A}}+\frac{m-2}{\ell_A+\ell_B}& -\frac{m/2-1}{\ell_A+\ell_B} & 0&\cdots&0&-\frac{m/2-1}{\ell_A+\ell_B}&-\frac{1}{\ell_A}\vspace{2pt}\\
-\frac{1}{\ell_B}& -\frac{m/2-1}{\ell_A+\ell_B}&\frac{m}{\ell_{C_3}} +\frac{2}{\ell_{B}}+\frac{m-2}{\ell_A+\ell_B}& -\frac{m/2-1}{\ell_A+\ell_B}&0&\cdots&0&-\frac{1}{\ell_B}\vspace{2pt}\\
\vdots&&\ddots&&&\ddots&&\vdots\vspace{2pt}\\
-\frac{1}{\ell_A}&0&\cdots&0& -\frac{m/2-1}{\ell_A+\ell_B}&\frac{m}{\ell_{C_1}} +\frac{2}{\ell_{A}}+\frac{m-2}{\ell_A+\ell_B}&  -\frac{m/2-1}{\ell_A+\ell_B}&-\frac{1}{\ell_A}\vspace{2pt}\\
-\frac{1}{\ell_B}& -\frac{m/2-1}{\ell_A+\ell_B}&0&\cdots&0& -\frac{m/2-1}{\ell_A+\ell_B}&\frac{m}{\ell_{C_3}} +\frac{2}{\ell_{B}}+\frac{m-2}{\ell_A+\ell_B}&-\frac{1}{\ell_B}\vspace{2pt}\\
0&-\frac{1}{\ell_A}&-\frac{1}{\ell_B}&-\frac{1}{\ell_A}&-\frac{1}{\ell_B}&\cdots&-\frac{1}{\ell_B}&\frac{n}{\ell_A}+\frac{n}{\ell_B}
\end{pmatrix}$}.
\label{eq:adjacent-matrix}
\ea

\smallskip

\noindent\textbf{Adjacent regions on a circle.\,} 
The geometry is shown in Figure~\ref{fig:adjacent-circle}, and the replica manifold is similar to that in Figure \ref{fig:adjacent-replica}. The corresponding matrix $M_{m,n}$ is given by
\ba
\scalebox{0.587}{$
\begin{pmatrix}
\frac{n}{\tanh\o\ell_A}+\frac{n}{\tanh\o\ell_B}& -\frac{1}{\sinh\o\ell_A} &-\frac{1}{\sinh\o\ell_B}& -\frac{1}{\sinh\o\ell_A}&\cdots&-\frac{1}{\sinh\o\ell_B}&0\vspace{5pt}\\
-\frac{1}{\sinh\o\ell_A}&\frac m {\tanh\o\ell_C}+\frac 2 {\tanh\o\ell_A}+\frac{m-2}{\tanh\o\ell_{AB}}& -\frac{m/2-1}{\sinh\o\ell_{AB}}-\frac{m/2}{\sinh\o\ell_C} & 0&0&-\frac{m/2-1}{\sinh\o\ell_{AB}}-\frac{m/2}{\sinh\o\ell_C}&-\frac{1}{\sinh\o\ell_A}\vspace{5pt}\\
-\frac{1}{\sinh\o\ell_B}&-\frac{m/2-1}{\sinh\o\ell_{AB}}-\frac{m/2}{\sinh\o\ell_C}&\ddots&\ddots&\ddots&0&-\frac{1}{\sinh\o\ell_B}\vspace{5pt}\\
\vdots&0&\ddots&\ddots&\ddots&&\vdots\vspace{5pt}\\
-\frac{1}{\sinh\o\ell_A}&0&\ddots&\ddots&\ddots& -\frac{m/2-1}{\sinh\o\ell_{AB}}-\frac{m/2}{\sinh\o\ell_C}&-\frac{1}{\sinh\o\ell_A}\vspace{5pt}\\
-\frac{1}{\sinh\o\ell_B}& -\frac{m/2-1}{\sinh\o\ell_{AB}}-\frac{m/2}{\sinh\o\ell_C}&0&& -\frac{m/2-1}{\sinh\o\ell_{AB}}-\frac{m/2}{\sinh\o\ell_C}&\frac m {\tanh\o\ell_C}+\frac 2 {\tanh\o\ell_B}+\frac{m-2}{\tanh\o\ell_{AB}}&-\frac{1}{\sinh\o\ell_B}\vspace{5pt}\\
0&-\frac{1}{\sinh\o\ell_A}&-\frac{1}{\sinh\o\ell_B}&-\frac{1}{\sinh\o\ell_A}&\cdots&-\frac{1}{\sinh\o\ell_B}&\frac{n}{\tanh\o\ell_A}+\frac{n}{\tanh\o\ell_B}
\end{pmatrix}$}.\label{eq:adjacent-circle_Mmn}
\ea

\subsection{Determinants}

A $2n\times 2n$ matrix ($n\ge2$) of the form
\be
\label{eq:m1}
M_1=
\begin{pmatrix}
a & -1 & 0&\cdots&0&-1\\
-1&b & -1&0&\cdots&0\\
0&-1&a & -1&\cdots&0\\
\vdots&&\vdots&&&\vdots\\
0&\cdots&0&-1&a& -1\\
-1&0&\cdots&0&-1&b 
\end{pmatrix},
\ee
with $ab>4$, has determinant
\be
\label{eq:det1}
\det M_1=2\cosh 2n\theta-2=4\(\sinh n\theta\)^2\, , \quad \mathrm{with} \quad \theta=\mathrm{arccosh} \frac{\sqrt{ab}}{2}\,.
\ee
The observation is that any term contributing to the determinant must have equal numbers of factors  of $a$ and $b$. 
This means we can replace all the diagonal terms by $\sqrt{ab}$
without affecting the determinant.
The new matrix becomes a circulant matrix
\be
M'_1=\mathrm{circ}\left\{ \sqrt{ab},-1,0,\cdots,0,-1\right\},
\ee
whose determinant has a formula
\be
\det M_1=\det M'_1=\prod_{j=1}^{2n}(\sqrt{ab}-e^{\frac{2\pi i j}{2n}} -e^{\frac{2\pi i (2n-1)j }{2n}})=\prod_{j=1}^{2n}\(-\sqrt{ab} +2\cos \frac{2\pi j}{2n}\) . 
\ee
To analytically continue the above expression to non-integer $n$, we use the property of Chebyshev polynomials of the first kind $T_{2n}$\,:
\be
\prod_{j=1}^{2n}\(2a +2\cos \frac{2\pi j}{2n}\)=2(T_{2n}(a)-1)\,,\quad \mathrm{with} \quad T_{2n}(\cos \theta)=\cos 2n\theta\,.
\ee
Then \eqref{eq:det1} follows immediately.

A $(2n+2)\times (2n+2)$ matrix of the form
\be
\label{eq:m3}
M_3=
\begin{pmatrix}
nx& \a & \b &\cdots& \b&0\\
\a&a &0&\cdots&0&\a\\
\b&0&b & \cdots&0&\b\\
\vdots&&\vdots&&&\vdots\\
\b&0&\cdots&0&b& \b\\
0&\a&\cdots&\a&\b&nx
\end{pmatrix},
\ee
has determinant
\be
\label{eq:det3}
\det M_3=n^2\(x^2a^n b^n-2xa^{n-1}b^{n-1}(b\a^2 +a \b^2)\)\,.
\ee
This can be shown by subtracting the first column by the last column and the first row by the last row.

A $(2n+2)\times (2n+2)$ matrix ($n\ge 2$) obtained by replacing the center diagonal matrix in $M_3$ by $M_1$
\be
\label{eq:m4}
M_4=
\begin{pmatrix}
nx&\a&\b&\a&\b&\cdots&\b&0\\
\a&a & -1 & 0&\cdots&0&-1&\a\\
\b&-1&b & -1&0&\cdots&0&\b\\
\a&0&-1&a & -1&\cdots&0&\a\\
\vdots&&\vdots&&&\vdots\\
\a&0&\cdots&0&-1&a& -1&\a\\
\b&-1&0&\cdots&0&-1&b &\b\\
0&\a&\b&\a&\b&\cdots&\b&nx
\end{pmatrix},
\ee
has determinant
\be
\label{eq:det4}
\det M_4=n^2\frac{\sinh^2n\theta}{\sinh^2\theta}\( (a b-4)x^2-2x\(\a^2 b+\b^2 a+4 \a\b \)\), \quad \mathrm{with}\quad \theta=\mathrm{arccosh} \frac{\sqrt{ab}}{2}\,.
\ee
This can be shown by calculating the minors of $M_1$, which are expressed in terms of Chebyshev polynomials of the second kind.

\section{Neumann boundary conditions}
\label{app:neu}

\begin{figure}[b]
\centering
\includegraphics[scale=1.1]{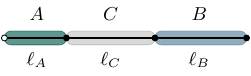}
\caption{Configuration with both Dirichlet and Neumann BCs. As before, we use hollow circle to denote Dirichlet BC and solid dots to denote free (Neumann) BC.}
\label{fig:adjacent-nd}
\end{figure}

Here we consider Neumann BCs at the boundaries of a finite system. We study the effect of changing one of the BCs from Dirichlet to Neumann, see, e.g., Figure~\ref{fig:adjacent-nd}. We shall show that imposing Neumann BC in the massless theory is equivalent to imposing Dirichlet BC and then letting the boundary go to infinity.

If we impose Neumann BC on a boundary, the boundary value $\phi_N$ of the field is an additional anchor to be integrated on in the calculation of partition functions. In fact this integral can be done easily. Suppose that in a replica graph, this boundary vertex is connected to another inner vertex with $\phi=\phi_i$ with an edge of length $\ell_N$, then the integral over $\phi_N$ gives
\be
\label{eq:int1}
\int d\phi_N  K(\phi_N,\phi_i;\ell_N)=1\,.
\ee
Thus, the final result does not depend on $\ell_N$. Now, a propagator that connects an inner vertex to a boundary vertex with Dirichlet BC contributes an exponential factor to the Gaussian integral, whose dependence on the length between the vertices disappears when sent to infinity, similarly as \eqref{eq:int1}. Hence, imposing Neumann BCs is equivalent to imposing Dirichlet BCs and letting the boundary go to infinity.

As a concrete example, let us compute the reflected entropy on a finite interval with Dirichlet-Neumann BCs with configuration shown in Figure~\ref{fig:adjacent-nd}. The replica manifold is similar to that in Figure~\ref{fig:disjoint-replica}, only with the BCs changed. Labelling the field values at these boundaries as $\phi_{2,i},\cdots,\phi_{2n,i}$ where $i=1,\cdots,m$, 
we now have
\ba
\Z_{m,n}&= \int d\phi_1\hspace{-1pt}\scalebox{0.9}{$\cdots$} d\phi_{2n}d\phi_{2,1}\hspace{-1pt}\scalebox{0.9}{$\cdots$}d\phi_{2,m}\hspace{-1pt}\scalebox{0.9}{$\cdots$} d\phi_{2n,1}\hspace{-1pt}\scalebox{0.9}{$\cdots$} d\phi_{2n,m} \,K(\phi_{2,1},\phi_2;\ell_B)\cdots K(\phi_{2,m},\phi_2;\ell_B)\cdots \nn\\
&\hspace{2.1cm}\times K(0,\phi_1;\ell_A)^m K(\phi_1,\phi_2;\ell_C)^{m/2} K(0,\phi_3;\ell_A)^{m} K(\phi_2,\phi_3;\ell_C)^{m/2} \cdots\nn\\
&=\int d\phi_1\cdots d\phi_{2n} K(0,\phi_1;\ell_A)^m K(\phi_1,\phi_2;\ell_C)^{m/2} K(0,\phi_3;\ell_A)^{m} K(\phi_2,\phi_3;\ell_C)^{m/2}\cdots\nn\\
&= \pi^{(1-m)n}(\ell_A\ell_C)^{-mn/2}\( \det M_{m,n}\)^{-1/2}, \label{eq:nb-partition}
\ea
where
\be
M_{m,n}=
\begin{pmatrix}
\frac{m}{\ell_A}+\frac{m}{\ell_C} & -\frac{m/2}{\ell_C} & 0&\cdots&0&-\frac{m/2}{\ell_C}\\
-\frac{m/2}{\ell_C}&\frac{m}{\ell_C} & -\frac{m/2}{\ell_C}&0&\cdots&0\\
0&-\frac{m/2}{\ell_C}&\frac{m}{\ell_A}+\frac{m}{\ell_C} & -\frac{m/2}{\ell_C}&\ddots&\vdots\\
\vdots&\ddots&\ddots&\ddots&\ddots&0\\
0&\cdots&0&-\frac{m/2}{\ell_C}&\frac{m}{\ell_A}+\frac{m}{\ell_C} & -\frac{m/2}{\ell_C}\\
-\frac{m/2}{\ell_C}&0&\cdots&0&-\frac{m/2}{\ell_C}&\frac{m}{\ell_C} 
\end{pmatrix}.
\label{eq:nb-matrix}
\ee
One clearly sees that the final result boils down to setting $\ell_B\rightarrow\infty$ in Section \ref{sec:disjoint}.

For the massive case, the equivalence of Neumann BC with Dirchlet at infinity does not carry through, since \eqref{eq:int1} does not hold. However, direct calculation with Neumann BC can be done as well.

\section{Deflected entropy}\label{deflected}

Recently introduced in \cite{Dutta:2022kge}, the deflected entropy is defined as a ``modular flowed'' version of reflected entropy. It is the entanglement entropy of the following deflected density matrix
\be
\rho^{(\sigma)}_{AA^*}=\tr_{BB^*}\ket{\rho_{AB}^{1/2+i \sigma}}\bra{\rho_{AB}^{1/2+i \sigma}}.
\ee
Similarly as the reflected entropy, the deflected entropy can also be calculated using a replica trick \cite{Dutta:2022kge} by considering
\be
\rho^{(m,\rho)}_{AA^*}=\tr_{BB^*}\ket{\rho_{AB}^{m/2+\rho}}\bra{\rho_{AB}^{m/2-\rho}},
\ee
and then performing the analytic continuation $\rho=i\sigma$. The minus sign comes from Hermitian conjugation.

We now compute the deflected entropy of two disjoint regions, as depicted in Figure \scref{fig:disjoint-l}, for massless Lifshitz theory. The replica graph is almost the same as Figure \ref{fig:disjoint-replica}, with the difference that now we have alternating numbers ($m/2-\rho$ and $m/2+\rho$) of length $\ell_C$ edges connecting two adjacent vertices. The matrix \eqref{eq:disjoint-matrix} for the deflected entropies reads
\be
M_{m,n}=
\begin{pmatrix}
\frac{m}{\ell_A}+\frac{m}{\ell_C} & -\frac{m/2+\rho}{\ell_C} & 0&\cdots&0&-\frac{m/2-\rho}{\ell_C}\\
-\frac{m/2+\rho}{\ell_C}&\frac{m}{\ell_B}+\frac{m}{\ell_C} & -\frac{m/2-\rho}{\ell_C}&0&\cdots&0\\
0&-\frac{m/2-\rho}{\ell_C}&\frac{m}{\ell_A}+\frac{m}{\ell_C} & -\frac{m/2+\rho}{\ell_C}&\ddots&\vdots\\
\vdots&\ddots&\ddots&\ddots&\ddots&0\\
0&\cdots&0&-\frac{m/2-\rho}{\ell_C}&\frac{m}{\ell_A}+\frac{m}{\ell_C} & -\frac{m/2+\rho}{\ell_C}\\
-\frac{m/2-\rho}{\ell_C}&0&\cdots&0&-\frac{m/2+\rho}{\ell_C}&\frac{m}{\ell_B}+\frac{m}{\ell_C} 
\end{pmatrix},
\label{eq:deflected-matrix}
\ee
whose determinant can be calculated using recursion.
The $(m,\sigma)$-deflected entropy is of the same form as \eqref{eq:re}, i.e.
\be
S^D_{m,\sigma}(A:B)=\frac{1}{\sqrt{1-\eta_{m,\sigma}}}\log\(\frac{1+\sqrt{1-\eta_{m,\sigma}}}{\sqrt{\eta_{m,\sigma}}}\)  -\log \(2\sqrt{\eta_{m,\sigma}^{-1}-1}\)\hspace{-2pt},
\ee
with 
\be
\eta_{m,\sigma}=\frac{1+4 \sigma^2/m^2}{\eta^{-1}+4 \sigma^2/m^2}\,,
\ee
and $\eta=\frac{\ell_A\ell_B}{\ell_{AC}\ell_{BC}}$ is the cross-ratio. The deflected entropy is then obtained by setting $m=1$\,.

\end{document}